\documentclass{emulateapj}

\slugcomment{Last revision 29-Nov-2003}

\shortauthors{Sheldon et al.}
\shorttitle{Galaxy-mass Correlations in SDSS}                                  

\newcommand{\mpname}[1]{#1_color.eps}
\newcommand{\clraitoff}{red}
\newcommand{\lumblack}{(black)}
\newcommand{\lumblue}{(blue)}
\newcommand{\lumred}{(red)}
\newcommand{\vdisred}{(red-dashed curve)}
\newcommand{\vdisblue}{(blue-solid curve)}

\newcommand{\umag}{$u$}
\newcommand{\gmag}{$g$}
\newcommand{\rmag}{$r$}
\newcommand{\imag}{$i$}
\newcommand{\zmag}{$z$}
\newcommand{\gmr}{$g-r$}
\newcommand{\deltasig}{$\Delta \Sigma$}

\newcommand{\deltacross}{$\Delta \Sigma_\times$}

\newcommand{\photo}{\texttt{PHOTO}}

\newcommand{\spectroone}{\texttt{SPECTRO1d}}
\newcommand{\spectrotwo}{\texttt{SPECTRO2d}}

\newcommand{\eclass}{\texttt{ECLASS}}
\newcommand{\eclasscut}{-0.06}
\newcommand{\gmrcut}{0.7}

\newcommand{\hrs}{$^{\mathrm h}$}
\newcommand{\minutes}{$^{\mathrm m}$}

\newcommand{\ugriz}{$u, g, r, i, z$}
\newcommand{\polarization}{polarization}

\newcommand{\wgm}{$w_{gm}$}
\newcommand{\wgg}{$w_{gg}^p$}

\newcommand{\xigg}{$\xi_{gg}$}

\newcommand{\xigm}{$\xi_{gm}$}

\newcommand{\numspec}{127,001}
\newcommand{\numspecvlim}{10,277}
\newcommand{\numrand}{1,270,010}

\newcommand{\numvdis}{49,024}
\newcommand{\numsource}{9,020,388}

\clearpage

\begin{document}

\title{The Galaxy-mass Correlation Function Measured from Weak Lensing in the 
SDSS}

\author{
Erin S. Sheldon,\altaffilmark{1,2}
David E. Johnston,\altaffilmark{1,2}
Joshua A. Frieman,\altaffilmark{1,2,3}
Ryan Scranton,\altaffilmark{4}
Timothy A. McKay,\altaffilmark{5}
A. J. Connolly,\altaffilmark{4}
Tam\'as Budav\'ari,\altaffilmark{6,7}
Idit Zehavi,\altaffilmark{8}
Neta A. Bahcall,\altaffilmark{9}
J. Brinkmann,\altaffilmark{10}
and Masataka Fukugita\altaffilmark{11}
}

\altaffiltext{1}{Center for Cosmological Physics, The University of Chicago, 5640 South Ellis Avenue Chicago, IL  60637}\altaffiltext{2}{Department of Astronomy and Astrophysics, The University of Chicago, 5640 South Ellis Avenue, Chicago, IL 60637.}
\altaffiltext{3}{Fermi National Accelerator Laboratory, P.O. Box 500, Batavia, IL 60510.}
\altaffiltext{4}{Department of Physics and Astronomy, University of Pittsburgh, 3941 O'Hara Street, Pittsburgh, PA 15260.}
\altaffiltext{5}{Department of Physics, University of Michigan, 500 East University, Ann Arbor, MI 48109-1120.}
\altaffiltext{6}{Department of Physics of Complex Systems, E\"otv\"os University, Budapest, Pf.\ 32, H-1518 Budapest, Hungary.}
\altaffiltext{7}{Department of Physics and Astronomy, The Johns Hopkins University, 3400 North Charles Street, Baltimore, MD 21218-2686.}
\altaffiltext{8}{Steward Observatory, University of Arizona, 933 North Cherry Avenue, Tucson, AZ 85721.}
\altaffiltext{9}{Princeton University Observatory, Peyton Hall, Princeton, NJ 08544.}
\altaffiltext{10}{Apache Point Observatory, P.O. Box 59, Sunspot, NM 88349.}
\altaffiltext{11}{Institute for Cosmic Ray Research, University of Tokyo, 5-1-5 Kashiwa, Kashiwa City, Chiba 277-8582, Japan.}

\begin{abstract}
We present galaxy-galaxy lensing measurements over scales 0.025 to 10 $h^{-1}$
Mpc in the Sloan Digital Sky Survey. Using a flux-limited sample of \numspec\
lens galaxies with spectroscopic redshifts and mean luminosity $\langle
L\rangle \sim L_*$ and \numsource\ source galaxies with photometric redshifts,
we invert the lensing signal to obtain the galaxy-mass correlation function
$\xi_{gm}$. We find $\xi_{gm}$ is consistent with a power-law, $\xi_{gm} =
(r/r_0)^{-\gamma}$, with best-fit parameters $\gamma = 1.79 \pm 0.06$ and $r_0
= (5.4 \pm 0.7)(0.27/\Omega_m)^{1/\gamma} h^{-1}$ Mpc.  At fixed separation,
the ratio $\xi_{gg}/\xi_{gm} = b/r$ where $b$ is the bias and $r$ is the
correlation coefficient.  Comparing to the galaxy auto-correlation function for
a similarly selected sample of SDSS galaxies, we find that $b/r$ is
approximately scale independent over scales $0.2-6.7 h^{-1}$ Mpc, with mean
$\langle b/r \rangle = (1.3 \pm 0.2)(\Omega_m/0.27)$.  We also find no scale
dependence in $b/r$ for a volume limited sample of luminous galaxies $(-23.0 <
M_r < -21.5)$.  The mean $b/r$ for this sample is $\langle b/r \rangle_{Vlim} =
(2.0 \pm 0.7)(\Omega_m/0.27)$.  We split the lens galaxy sample into subsets
based on luminosity, color, spectral type, and velocity dispersion, and see
clear trends of the lensing signal with each of these parameters. The amplitude
and logarithmic slope of $\xi_{gm}$ increases with galaxy luminosity.  For high
luminosities ($L \sim 5 L_*$), \xigm\ deviates significantly from a power
law. These trends with luminosity also appear in the subsample of red galaxies,
which are more strongly clustered than blue galaxies.
\end{abstract}




\section{Introduction} \label{intro}

The measurement of galaxy clustering has long been a primary tool in
constraining structure formation models and cosmology. Yet the power of galaxy
surveys to discriminate between models is partially compromised by the fact
that they provide only an indirect measure of the underlying mass distribution,
subject to considerable uncertainties in the {\it bias}, that is, in how
luminous galaxies trace the mass. In this context, the galaxy-mass
cross-correlation function $\xi_{gm}$ can provide important additional
information, since it is in some sense a `step closer' to the clustering of
mass. Moreover, comparing $\xi_{gm}$ with the galaxy auto-correlation function
$\xi_{gg}$ yields a measure of the bias and therefore a constraint on theories
of galaxy formation. In this paper we measure the correlation between galaxies
and mass using weak gravitational lensing.

In the current paradigm of structure formation, the formation of galaxies is
heuristically divided into two parts. On large scales, cosmological parameters
and the properties of the dark matter determine the growth of density
perturbations and the eventual formation of massive dark halos. On smaller
scales, hydrodynamic and other processes shape how luminous galaxies form
within dark matter halos and how they evolve as halos accrete and merge. A
natural consequence of this picture is that the galaxy distribution is related
to but differs in detail from the mass distribution. This difference arises in
part because halos are more strongly clustered than the dark matter as a whole,
and more massive halos are more strongly clustered than less massive ones
\citep{Kaiser84}. In addition, the efficiency of forming luminous galaxies of
different types and luminosities varies (primarily) with halo mass.

The galaxy bias parameter can be defined in a number of ways, but a traditional
one is as the ratio of the galaxy and mass auto-correlation functions at fixed
separation \citep{Kaiser84},
\begin{equation} \label{eq:bias_def}
b^2 = \frac{\xi_{gg}}{\xi_{mm}} ~~.
\end{equation}
The amplitude of the galaxy-mass cross-correlation $\xi_{gm}$ relative to
$\xi_{gg}$ and $\xi_{mm}$ can be expressed in terms of the correlation
coefficient \citep{Pen98},
\begin{equation} \label{eq:r_def}
r = \frac{\xi_{gm}}{(\xi_{mm}\xi_{gg})^{1/2}} ~,
\end{equation}
so that $\xi_{gm} = b r \xi_{mm}$.  In general, $b$ and
$r$ can be time-dependent functions of the pair separation and depend on galaxy
properties.  Since we will be comparing the galaxy-galaxy and galaxy-mass
correlation functions, we will constrain the ratio
\begin{equation} \label{bgmdef}
\frac{\xi_{gg}}{\xi_{gm}} = \frac{b}{r}~.
\end{equation}

Galaxy surveys have provided a wealth of information on the 
behavior of the galaxy bias $b$ as a function of galaxy luminosity 
and type. For example, on scales $r < 20 h^{-1}$ Mpc, the clustering 
amplitude \xigg\ increases with luminosity 
\citep{Norberg01,Zehavi02,Norberg02,Zehavi04}, 
while the amplitude and shape of $\xi_{gg}$ 
vary systematically from early to late galaxy types 
\citep{Davis76,Norberg01,Zehavi02}. Moreover, 
the nearly power-law behavior of \xigg\, as well 
as the small departures therefrom \citep{Zehavi03}, combined with 
the assumption that the dark matter distribution is 
described by cold dark matter models, indicate that the bias is 
scale-dependent on these scales.

On larger scales, $r > 20 h^{-1}$ Mpc, there is evidence from higher-order
galaxy correlations \citep{Frieman99,Szapudi02,Verde02}, from the cosmic shear
weak lensing power spectrum \citep{Hoekstra02c,Jarvis03}, and from comparison
of the 2dF galaxy power spectrum \citep{Percival01} with the WMAP cosmic
microwave background (CMB) temperature angular power spectrum \citep{Spergel03}
that the linear galaxy bias parameter $b_{lin}$ is of order unity for optically
selected $L_*$ galaxies.  (Here, $b_{lin}^2$ is the ratio of the galaxy
correlation function to the mass correlation function computed in linear
perturbation theory; it is related but not identical to the bias defined in
Equation~\ref{eq:bias_def}.)  Measurement of the parameter $\beta =
\Omega_m^{0.6}/b_{lin}$ from redshift space distortions in galaxy surveys,
combined with independent evidence that $\Omega_m \simeq 0.3$, also indicates
$b_{lin}(L_*) \simeq 1$ on large scales \citep{PeacockNature2001}.

In recent years, weak gravitational lensing has become a powerful tool for
probing the distribution and clustering of mass in the Universe. We focus on
galaxy-galaxy lensing, the distortion induced in the images of background
(source) galaxies by foreground lens galaxies. Although the typical distortion
induced by a galaxy lens is tiny ($\sim 10^{-3}$) compared to the intrinsic
ellipticities of the source galaxies ($\sim 0.3$), the signals from a large
sample of lens galaxies can be stacked, providing a mean measurement with high
signal to noise.  The mean lensing signal can be used to infer the galaxy-mass
cross-correlation function which, when compared with the galaxy
auto-correlation function, constrains the amplitude and scale dependence of the
bias.

The first detection of galaxy-galaxy lensing was made by \citet{Brainerd96},
and the field has progressed rapidly since then
\citep{Dell96,Griffiths96,Hudson98,fis00,Wilson01,Smith01,Mckay02,Hoekstra03a}.
The first high S/N measurements were made in the Sloan Digital Sky Survey
(SDSS) \citep{fis00}.  Recent studies have benefited from improved data
analysis and reduction techniques and from surveys which are specifically
designed for lensing \citep{Hoekstra01b}.  Most work in galaxy lensing has
concentrated on its power to constrain galaxy halo parameters
\citep{Brainerd96,Hudson98,fis00,Hoekstra03a}.  However, galaxy lensing has
also been used to measure the bias directly \citep{Hoekstra01b,Hoekstra02b};
their results indicate that $b$ and $r$ are scale-dependent over scales $\sim
0.1-5 h^{-1}$ Mpc, but that the ratio $b/r$ is nearly constant at $b/r \simeq
1.1$ over this range.

A significant step forward in galaxy-galaxy lensing came with the use of
samples of lens galaxies with spectroscopic redshifts \citep{Smith01,Mckay02}.
Lensing measurements could then be made as a function of physical rather than
angular separation, placing lensing correlation measurements on a par with the
auto-correlation measurements from galaxy redshift surveys. Incorporation of
photometric redshifts for the source galaxies \citep{Hudson98} also
substantially reduces errors in the lens mass calibration due to the breadth of
the source galaxy redshift distribution.

In this paper, we study galaxy-mass correlations in the SDSS using weak
gravitational lensing.  Using a sample of \numspec\ galaxies with spectroscopic
redshifts and \numsource\ galaxies with photometric redshifts, we measure the
lensing signal with high S/N over scales from $0.025-10 h^{-1}$ Mpc.  This is
the first galaxy lensing study to incorporate both spectroscopic lens redshifts
and photometric source redshifts, it is by far the largest galaxy lens-source
sample compiled, and it extends to scales larger than previous galaxy-galaxy
measurements. A similar spectroscopic sample has been used for galaxy
auto-correlation measurements in the SDSS \citep{Zehavi03}.  We compare the
galaxy-mass and galaxy-galaxy correlations to constrain $b/r$ over scales
from 200 kpc to 10 Mpc.  We also use the spectroscopic and photometric data
from the SDSS to divide the lens galaxy sample by luminosity, color, spectral
type, and velocity dispersion.  We see clear dependences of $\xi_{gm}$ on each
of these properties.

The layout of the paper is as follows: In \S \ref{lensing} we introduce lensing
and the measurement methods.  In \S \ref{data} we discuss the SDSS data,
reductions, and sample selection.  The basic measurement of the lensing signal
\deltasig\ is presented in \S \ref{gglmeas:deltasig}, and important checks and
corrections for systematic errors using random points are discussed in \S
\ref{gglmeas:random}. In \S \ref{gglmeas:xigm}-\ref{gglmeas:vlim} we use the
data to infer the galaxy-mass correlation function $\xi_{gm}$ and compare it
with independent measurements of $\xi_{gg}$ to constrain the bias.  In \S
\ref{gmcflum}-\ref{gmcftype} we explore the dependence of galaxy-mass
correlations on galaxy luminosity and type.  We conclude in \S \ref{discussion}
and discuss possible systematic errors in the Appendix.

Throughout this paper, where necessary we use a Friedman-Robertson-Walker
cosmology with $\Omega_M$ = 0.27, $\Omega_{\Lambda}$ = 0.73, and H$_0$ = 100 h
km/s.  All distances, densities, and luminosities are expressed in comoving
coordinates.

\section{Lensing and Galaxy-mass Correlations} \label{lensing}

\subsection{Gravitational Shear and the Galaxy-mass Correlation Function} 
\label{lensing:shear}

In this section we review the relation between the induced shear, which can be
estimated from source galaxy shape measurements, the galaxy-mass cross
correlation function $\xi_{gm}$, and the projected cross-correlation function
\wgm.  The tangential shear, $\gamma_T$, azimuthally averaged over a thin
annulus at projected radius $R$ from a lens galaxy, is directly related to the
projected surface mass density of the lens within the aperture,
\begin{equation} \label{eq:gammat}
\gamma_T \times \Sigma_{crit} =  \overline{\Sigma}(< R) -\overline{\Sigma}(R)
\equiv \Delta \Sigma ~,
\end{equation}
where $\overline{\Sigma}(< R)$ is the mean surface density within radius $R$,
and $\overline{\Sigma}(R)$ is the azimuthally averaged surface density at
radius $R$ \citep{Escude91,Kaiser94,Wilson01}.  The proportionality constant
$\Sigma_{crit}$ encodes the geometry of the lens-source system,
\begin{equation} \label{eq:sigmacrit}
\Sigma_{crit}^{-1} = \frac{4 \pi G D_{LS} D_L}{c^2 D_S}~,
\end{equation}
where $D_{L}$, $D_S$, and $D_{LS}$ 
are angular diameter distances to lens, source, and between
lens and source.  

Due to the subtraction in equation \ref{eq:gammat}, uniform mass sheets (such
as the mean density of the universe $\overline{\rho} = \Omega_m\rho_{crit}$) do
not contribute to \deltasig---it measures the mean {\it excess} projected mass
density.  The mean excess mass density at radius $r$ from a galaxy is
$\overline{\rho}~\xi_{gm}(r)$. The mean excess projected density $\Sigma(R)$ is
given by the radial integral:
\begin{equation}
\langle \Sigma(R) \rangle = 
      \int \overline{\rho} \xi_{gm}(x,y,z) dz \equiv \overline{\rho} w_{gm}(R) ~, 
\end{equation}
where \wgm\ is the projected galaxy-mass correlation function and $R = (x^2 +
y^2)^{1/2}$ is the projected radius.  The observable \deltasig\ is itself an
integral over $\Sigma(R)$ and hence \wgm:
\begin{equation} \label{eq:deltasig_and_wgm}
\langle \Delta\Sigma(R) \rangle = \overline{\rho} \times \left[ \frac{2}{R^2} 
\int_0^R R^\prime dR^\prime ~w_{gm}(R^\prime) -w_{gm}(R) \right]
\end{equation}

If the cross-correlation function can be approximated by a power-law in
separation, $\xi_{gm} = (r/r_0)^{-\gamma}$, then \wgm\ can be written as
\begin{equation} \label{xigm_def}
w_{gm}(R)  =  F(\gamma, r_0) R^{1-\gamma} ~,
\end{equation}
where $ F(\gamma, r_0) = r_0^{\gamma}
\Gamma(0.5)\Gamma[0.5(\gamma-1)]/\Gamma(0.5\gamma)$ \citep{DavisPeebles83}.  In
that case, the mean lensing signal \deltasig\ is also a power law with index
$\gamma - 1$ and is simply proportional to $\overline{\rho}w_{gm}$,
\begin{equation} \label{deltasig2wgm}
\langle \Delta \Sigma (R)\rangle = \left(
\frac{\gamma -1}{3-\gamma} \right)\overline{\rho} w_{gm}(R)~.
\end{equation}


More generally, the three-dimensional galaxy-mass correlation function can be
obtained by inverting \deltasig\ directly.  Differentiating equation
\ref{eq:deltasig_and_wgm}, we find
\begin{equation}
-\overline{\rho} \frac{d w_{gm}}{d R} = \frac{d \Delta \Sigma}{d R} + 2 \frac{\Delta \Sigma}{R}.
\end{equation}
The derivative $d w_{gm}/ d R$ can be integrated to obtain \xigm\ using an
Abell formula \citep{Saunders92}:
\begin{equation} \label{eq:saunders}
\xi_{gm}(r) =  \frac{1}{\pi} \int_r^\infty dR \frac{-dw_{gm}/dR}{(R^2 - r^2)^{1/2}}
\end{equation}
In practice, the data only cover a finite range of scales up to $r=R_{max}$.
The estimated \xigm\ integrating to $R_{max}$ is related to the true \xigm\ by
\begin{equation}
\xi_{gm}^{est}(r) = \xi_{gm}(r) - \frac{1}{\pi} \int_{R_{max}}^{\infty} dR \frac{-dw_{gm}/dR}{(R^2 - r^2)^{1/2}}
\end{equation}
where the last term reminds us of the (unknown) contribution from scales beyond
those for which we have measurements.
Provided the integrand falls sufficiently fast with separation, this term is
negligible for scales $r$ smaller than a fraction of $R_{max}$.  Furthermore,
since $\xi_{gm}$ is linear in $\Delta \Sigma$, the covariance matrix of the
latter can be straightforwardly propagated to that of the former.

\subsection{Estimating \deltasig}\label{lensing:deltasig}

We estimate the shear by measuring the tangential component of the source
galaxy ellipticity relative to the the lens center, $e_+$, also known as the
E-mode.  In general, the shear is related in a complex way to $e_+$
\citep{SchneiderSeitz95}, but in the weak lensing regime the relationship is
linear:
\begin{equation} \label{eq:ellip_induce}
e_+ = 2 \gamma_T \mathcal{R} + e_+^{int} ~,
\end{equation} 
where $e_+^{int}$ is the intrinsic ellipticity of the source, $\gamma_T$ is the
shear, and $\mathcal{R}$ is the ``responsivity'' (see equation
\ref{eq:responsivity}).  The assumption behind weak lensing measurements is
that the source galaxies are randomly oriented in the absence of lensing, in
which case their intrinsic shapes constitute a large but random source of error
on the shear measurement.  This ``shape noise'' is the dominant source of noise
for most weak lensing measurements.  We discuss limits on intrinsic
correlations between galaxy ellipticities in the appendix.

The other component of the ellipticity, $e_\times$, also known as the B-mode,
is measured at 45\degr\ with respect to the tangent. The average B-mode should
be zero if the induced shear is due only to gravitational lensing
\citep{Kaiser95,Luppino97}.  This provides an important test for systematic
errors, such as uncorrected PSF smearing, since they generally contribute to
both the E- and B-modes.

In order to estimate \deltasig\ from the shear, we must know the angular
diameter distances $D_L$, $D_S$, $D_{LS}$ for each lens-source pair (see
equations \ref{eq:gammat} and \ref{eq:sigmacrit}).  In the SDSS, we have
spectroscopic redshifts for all the lens galaxies, so that $D_L$ is measured to
high precision (assuming a cosmological model).  For the source galaxies, we
have photometric redshift estimates (photoz), with typical relative errors of
20-30\% (see \S \ref{data:imaging:photoz}), so there is comparable uncertainty
in the value of $\Sigma_{crit}$ for each lens-source pair.

Given the known redshifts of the lenses, the distribution of
errors in the source galaxy 
ellipticity, and the distribution of errors in the photometric
redshift for each source, we can write the likelihood for $\Delta \Sigma$ from 
all lens-source pairs,
\begin{equation} \label{eq:deltasig_like}
\mathcal{L}(\Delta\Sigma) = \prod_{j=1}^{N_{Lens}} \prod_{i=1}^{N_{Source}}
\int dz_s^i P(z_s^i) P(\gamma_T^i | z_s^i, z_L^j) ~,
\end{equation}
where $\gamma_T^i = e_+^i/2 \mathcal{R}$ is the shear estimator for the $i$th
source galaxy, $P(z_s^i)$ is the probability distribution for its redshift (the
product of the Gaussian error distribution returned by the photoz estimator and
a prior based on the redshift distribution for the source population; see \S
\ref{data:imaging:photoz}), and $P(\gamma_T^i | z_s^i, z_L^j)$ is the
probability distribution of the shear given the source and lens redshifts,
which is a function of the desired quantity \deltasig:
\begin{eqnarray} \label{eq:pgamma}
\lefteqn{P(\gamma_T^i | z_s^i, z_L^j) \varpropto}   \nonumber \\
& &
\exp
\left(-\frac{1}{2} 
\left[
\frac{\gamma_T^i - \Delta\Sigma \times \Sigma_{crit}^{-1} (z_s^i, z_L^j)}{\sigma(\gamma_T^i)}.
\right]^2
\right) 
\end{eqnarray}
In equation \ref{eq:pgamma}, $4\times\sigma^2(\gamma_T^i) = \sigma^2(e_+^i) +
\sigma^2_{SN}$: the shear uncertainty is the sum of the measurement variance
$\sigma^2(e_+^i)$ and the intrinsic variance in the shapes of the source
galaxies $\sigma^2_{SN}=\langle (e_+^{int})^2\rangle$. The shape noise measured
from bright, well-resolved galaxies is $\sigma_{SN} \approx 0.32$, and the
typical measurement error $\sigma(e_+)$ ranges from $\sim$0.05 for \rmag=18 to
$\sim$0.4 for \rmag=21.5.  The intrinsic shape distribution is not Gaussian as
we have assumed in equation \ref{eq:pgamma}, but it is symmetric.  Monte Carlo
simulations indicate that this approximation does not bias the measurement of
$\Delta \Sigma$ within our measurement uncertainties, provided we take
$\sigma_{SN}$ as the standard deviation of the non-Gaussian shape distribution.

Although the typical uncertainty in the source galaxy photometric redshifts is
20-30\%, this is small compared to the relative shear noise, which is typically
$\sigma_{SN}/\gamma_T \sim 300$\%.  Assuming shape error is the dominant source
of noise, we can approximate equation \ref{eq:deltasig_like} as
\begin{equation}
\log \mathcal{L}(\Delta\Sigma) = 
\sum_{j,i}
\left(-\frac{1}{2} 
\left[
\frac{\gamma_T^i - \Delta\Sigma \times \langle \Sigma_{crit}^{-1}\rangle_{j,i}}{\sigma(\gamma_T^i)}
\right]^2
\right)
\end{equation}
where we now use the critical surface density averaged over the photoz
distribution for each source galaxy,
\begin{equation} \label{eq:meansigcrit}
\langle \Sigma_{crit}^{-1} \rangle_{j,i} = \int dz_s^i P(z_s^i) \Sigma_{crit}^{-1}(z_s^i, z_L^j)
\end{equation}
Monte Carlo simulations indicate that the true likelihood approaches this
Gaussian approximation after stacking only a few hundred lenses.

The maximum likelihood solution is the standard weighted average,
\begin{equation}
\Delta\Sigma =  
{\sum_{j=1}^{N_{Lens}} \sum_{i=1}^{N_{Source}} \Delta\Sigma_{j,i} w_{j,i}
\over \sum_{j=1}^{N_{Lens}} \sum_{i=1}^{N_{Source}}w_{j,i}} ~,
\end{equation}
where  
\begin{eqnarray} \label{eq:dsigterms}
\Delta\Sigma_{j,i} & = & \gamma_T^i/\langle \Sigma_{crit}^{-1}\rangle_{j,i} \nonumber \\
w_{j,i} & = & \sigma_{j,i}^{-2} \nonumber \\
\sigma_{j,i} & = & \sigma(\gamma_T^i)/\langle \Sigma_{crit}^{-1}\rangle_{j,i}.
\end{eqnarray}  
Although this simple inverse variance weighting is not optimal \citep{Bern02},
it does lead to unbiased results and only a slight increase in the variance
of \deltasig.

As indicated by equation \ref{eq:ellip_induce}, the ellipticity induced by a
shear depends on the object's shear responsivity $\mathcal{R}$.  A measure of
how an applied shear alters the shape of a source, $\mathcal{R}$ depends on the
object's intrinsic ellipticity and is similar to the shear polarizability of
\citet{ksb95}. Following \citet{Bern02}, we calculate a mean responsivity as a
weighted average over all tangential ellipticities.
\begin{equation}
\mathcal{R} = \frac{ \sum_{j,i} w_{j,i} \left[ 1 - k_0 - k_1 (e_+^i)^2\right] }{\sum_i w_{j,i}}~,
\label{eq:responsivity}
\end{equation}
where the weights are the same as in equation \ref{eq:dsigterms}. We have again
assumed a Gaussian distribution of ellipticities so that $k_0$ and $k_1$ are
simple,
\begin{eqnarray}
k_0 & = & (1-f) \sigma^2_{SN},~~~k_1 = f^2, \nonumber   \\
f & = & \frac{\sigma^2_{SN}}{\sigma^2_{SN} + \sigma^2(e_+^i)}
\end{eqnarray}
The mean $\mathcal{R}$ for our sources is 0.86.

\section{Data} \label{data}

The Sloan Digital Sky Survey (SDSS; \cite{York00}) is an ongoing project to map
nearly 1/4 of the sky in the northern Galactic cap (centered at 12\hrs
20\minutes, +32.8\degr).  Using a dedicated 2.5 meter telescope located at
Apache Point Observatory in New Mexico, the SDSS comprises a photometric survey
in 5 bandpasses (\ugriz; \citet{Fukugita96}) to \rmag\ $\sim$ 22 and a
spectroscopic survey of galaxies, luminous red galaxies, quasars, stars, and
other selected targets. In addition, the survey covers 3 long, thin stripes in
the southern Galactic hemisphere; the central southern stripe, covering $\sim
200$ square degrees, will be imaged many times, allowing time-domain studies as
well as a deeper co-added image.


We select our lens galaxies from the SDSS main galaxy spectroscopic sample.
For our background sources we use only well-resolved galaxies drawn from the
photometric survey with well-measured photometric redshifts. Each sample is
described in detail below.

\subsection{Imaging Data} \label{data:imaging}

Imaging data are acquired in time-delay-and-integrate (TDI) or drift scan mode.
An object passes across the camera \citep{Gunn98} at the same rate the CCDs
read out, which occurs continuously during the exposure.  The object crosses
each of the 5 SDSS filters in turn, resulting in nearly simultaneous images in
each bandpass.  In order for the object to pass directly down the CCD columns,
the distortion across the field of view must be exceedingly small.  This is
advantageous for lensing, since distortions in the optics cause a bias in
galaxy shapes (\S \ref{data:imaging:correct}).  The distortion in the optics of
the SDSS 2.5m telescope is negligible \citep{Stough02} and can be ignored.

The imaging data are reduced through various software pipelines, including the
photometric (\photo, \cite{LuptonADASS01}), astrometric \citep{Pier03}, and
calibration \citep{Hogg01,Smith02} pipelines, leading to calibrated lists of
detected objects.  The calibrated object lists are subsequently fed through
various target selection pipelines \citep{Eisenstein01,Strauss02,Richards02}
which select objects for spectroscopic followup.

The shape measurements discussed in \S \ref{data:imaging:meas} are implemented
in \photo\ (\texttt{v5\_3}), so we work directly with the calibrated object
lists.  These lists contain, among many other things, position (RA,DEC),
several measures of the flux, diagnostic flags for the processing, and moments
of the light distribution for each object and for the local Point Spread
Function (PSF) \citep{Stough02}.  We augment the parameters measured by \photo\
with the probability that each object is a galaxy (\S \ref{data:imaging:sg})
and with photometric redshifts.

\subsubsection{Shape Measurements} \label{data:imaging:meas}

Weak lensing measurements rely on the assumption that source galaxy shapes are
an unbiased, albeit rather noisy, measure of the shear induced by foreground
lenses.  Therefore high S/N shape measurements and accurate corrections for
bias are crucial for weak lensing measurements.  For both shape measurements
and corrections we use techniques described in \cite{Bern02} (hereafter BJ02).

We determine the apparent shapes of objects from their flux-weighted second
moments.
\begin{equation}
Q_{m,n} = \sum_{m,n} I_{m,n} W_{m,n} x_m x_n ,
\label{data:imaging:meas:mom}
\end{equation} 
where $I_{m,n}$ is the intensity at pixel $m,n$ and $W_{m,n}$ is an elliptical
Gaussian weight function, iteratively adapted to the shape and size of the
object.  Initial guesses for the size and position of the object are taken from
the Petrosian radius and \photo\ centroid.  Objects are removed if the
iteration does not converge or if the iterated centroid wanders too far from
the \photo\ centroid.  The shape is parametrized by the polarization, or
ellipticity, components, defined in terms of the second moments,
\begin{eqnarray}
e_1 & = & { {Q_{11} - Q_{22}} \over {Q_{11} + Q_{22}} }  \\
e_2 & = & { {2Q_{12}} \over {Q_{11} + Q_{22}} }.
\end{eqnarray} \label{data:imaging:meas:e1e2}
The \polarization\ is related directly to the shear via equation
\ref{eq:ellip_induce}.  

The errors in the moments are calculated from the photon noise under the
assumption that each measurement is sky noise limited, which is a good
approximation for the faint sources.  Because the same pixels are used for
measuring both components of the ellipticity, there is a small covariance
between the components of the ellipticity.  This covariance is also calculated
and properly transformed when rotating to the tangential frame for shear
measurements.

Because the natural coordinates for the SDSS are survey coordinates
$(\lambda,\eta)$ \citep{Stough02}, we rotate the ellipticities into that
coordinate system for the lensing measurements.  The full covariance matrix for
$(e_1,e_2)$ is used to transform the errors under rotations.  We combine the
shape measurements from the \gmag, \rmag, and \imag\ bandpasses using the
covariance matrices.  This increases the S/N, simplifies the analysis, and
reduces bandpass-dependent systematic effects.  The \umag\ and \zmag\ bands are
much less sensitive and would contribute little to the analysis.

\subsubsection{PSF Reconstruction} \label{data:imaging:psf}

An anisotropic PSF, caused by instrumental and atmospheric effects, smears and
alters the shapes of galaxies in a way which can mimic lensing. In addition,
the finite size of the PSF---the seeing---tends to circularize the image,
reducing the measured ellipticity.  In order to correct for these effects, one
should, in principle, determine the exact shape and size of the PSF at the
position of every source galaxy.  Imaging in drift scan mode produces images
that are long, thin stripes on the sky.  Because the PSF varies over time
(along the direction of the scan) as well as across the camera, it must be
tracked as a function of position in the overall image (see \S
\ref{data:imaging:correct}).

The photometric pipeline uses Karhunen-Lo\`{e}ve (KL) decomposition
\citep{Hot33,Kar47,Loeve48} to model the PSF \citep{LuptonADASS01}. The PSF is
modeled on a frame-by-frame basis, where frames are defined as $2048\times
1490$ pixel chunks composing the long SDSS image.  A set of bright, isolated
stars are chosen from $\pm 2$ frames around the central frame, with typically
$15-25$ stars per frame.  These stellar images $P_{(i)}(u,v)$ are used to form
a set of KL basis functions or \emph{eigenimages} $B_r(u,v)$, in terms of which
the images can be reconstructed by keeping the first $n$ terms in the
expansion,
\begin{equation}
P_{(i)}(u,v) = \sum_{r=0}^{n-1}a_{(i)}^r B_r(u,v)
\end{equation}
where $P_{(i)}$ denotes the $i^{th}$ star, and $u,v$ are 
pixel positions relative to the object center.

The spatial dependence of the coefficients $a_{(i)}^r$ are determined via a
polynomial fit,
\begin{equation}
a_{(i)}^r \approx \sum_{l=m=0}^{l+m \le N} b_{lm}^r x_{(i)}^l y_{(i)}^m
\end{equation}
where $x,y$ are the coordinates of the center of the $i^{th}$ star 
relative to the center of the frame, $N$
is the highest order in $x,y$ included in the expansion, and $b_{lm}^r$ are
determined from minimizing 
\begin{equation}
\chi^2 = \sum_i\left( P_{(i)}(u,v) - \sum_{r=0}^{n-1} a_{(i)}^r B_r(u,v)\right)
\end{equation}
Only the stars on $\pm 1/2$ a frame surrounding the given frame are 
used to determine the spatial variation, while stars from $\pm 2$ frames
are used to determine the KL basis functions.

In \photo, the number of terms used from the KL basis is usually $n=3$; the
order of the spatial fit is $N=2$ unless there are too few stars, in which case
the fit may be order 1 or even 0 (rare). To determine the coefficients
$b_{lm}^r$, a total of $n(N+1)(N+2)/2$ constraints are needed, which may seem
like too many for the typical number of available stars.  There are many pixels
in each star, however, so the number of spatial terms $(N+1)(N+2)/2 = 6$ (for
quadratic fits) should be compared with the number of available stars.

The PSF is reconstructed at the position of each object and its second
moments are measured; these are used in the analytic PSF correction scheme
described in \S \ref{data:imaging:correct}.

\subsubsection{Shape Corrections} \label{data:imaging:correct}

To correct galaxy shapes for the effects of PSF dilution and anisotropy, we use
the techniques of BJ02 with the modifications specified in \cite{Hirata02}.
Rather than a true deconvolution, this is an approximate analytic
technique. The PSF is modeled as a transformation of the preseeing shape; to
remove the effects of the PSF, the inverse transform must be calculated.  This
transformation is performed in shear space (or \polarization\ space):
\begin{equation} \label{eq:psfcorrect}
R * [ (-e_{PSF}) \oplus e ] = (-e_{PSF}) \oplus e_0
\end{equation}
where $e$ is the preseeing \polarization, $e_0$ is the measured \polarization,
$e_{PSF}$ is the \polarization\ of the PSF, and $R$ is the resolution
parameter, which is related to the polarizability of \citet{ksb95}.  The
operator $\oplus$ is the shear addition operator defined in BJ02.  If the shear
addition operator were a simple addition, then equation \ref{eq:psfcorrect}
would reduce to:
\begin{equation}
e = [e_0 - (1-R)e_{PSF}]/R
\end{equation}

For unweighted moments ($W_{m,n}=1$) or if the objects and PSF have Gaussian
surface brightness profiles, the formulae in BJ02 are exact, and the resolution
parameter in this case is just $R=1-(s_{PSF}/s_{obj})^2$, where $s$ is the
linear size of the object or PSF.  The light profiles of galaxies and of the
PSF differ significantly from Gaussian, however.  Furthermore, as noted in \S
\ref{data:imaging:meas}, we use a Gaussian radial weight function to optimize
the S/N of object shape measurements.  As a consequence, the resolution
parameter $R$ must be derived in an approximate way that accounts for both the
weight function and the non-Gaussianity of the light profiles.  For this work
we use a weighted fourth moment to correct for higher order effects as
discussed in BJ02 and \citet{fis00}.


\subsubsection{Star-Galaxy Separation} \label{data:imaging:sg}

To separate stars and galaxies cleanly at all magnitudes, we use the Bayesian
method discussed in \cite{Scranton02}.  The method makes use of the
concentration parameter, which can be calculated from parameters output by
\photo.  The concentration is the difference between the object's PSF magnitude
and exponential disk magnitude. The PSF magnitude is derived by fitting the
local PSF shape to the object's light profile. The only free parameter in this
fit is the overall flux. An exponential disk is also fit to the object, but in
that case the scale length is also a free parameter.  Thus, large objects have
more flux in the exponential than the PSF fit and correspondingly large
concentration, while stars have concentration around zero.

\begin{figure}[bp]
\plotone{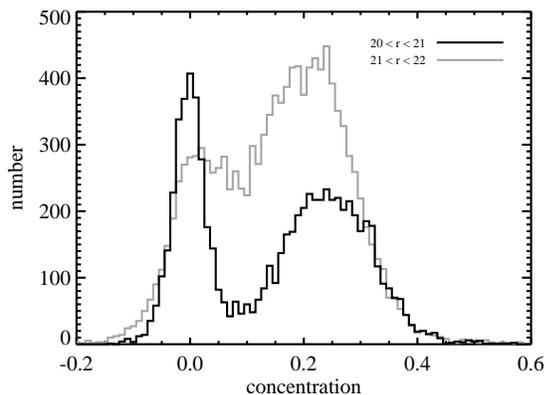} \figcaption{
Concentration distribution for objects
with $20 <~$\rmag$~< 21$ (dark curve) and $21 <~$\rmag$~< 22$ (light
curve). Stars have concentration near zero.  At faint magnitudes, stars and
galaxies are not as easily separated. \label{fig:conc} }
\end{figure}

At bright magnitudes, galaxies and stars separate cleanly in concentration
space. At fainter magnitudes, photometric errors increase and the distributions
overlap. This is demonstrated in figure \ref{fig:conc}, which shows the
distribution of concentration for objects with $20 <~$\rmag$~< 21$ and $21
<~$\rmag$~< 22$ drawn from 100 frames of a single SDSS imaging run (3325), with
mean seeing of 1.25\arcsec (typical of SDSS image quality).  In bad seeing
conditions, a smaller percentage of galaxies are larger than the PSF, again
making separation difficult.

If we know the distribution in concentration for galaxies and stars as a
function of seeing and magnitude, we can assign each object a probability $P_g$
that it is a galaxy, given its concentration, magnitude, and the local seeing.
For our source galaxy sample, we can then select objects which have high values
of $P_g$.  To map out the distribution in concentration, we use regions from
the SDSS Southern Survey which have been imaged many times.  Some of these
regions have been imaged as many as 16 times, with an average of about 8. By
averaging the flux for each object from the multiple exposures to obtain higher
S/N, a clean separation between the star and galaxy concentration distributions
is achieved to fainter magnitudes.  We have maps of the concentration
distribution for $16 <~$\rmag$~< 22$ and $0.9 <~$seeing$~< 1.8$, allowing us to
accurately calibrate $P_g$ and therefore define a clean galaxy sample.  In the
few regions of very bad seeing, we extrapolate the concentration
conservatively, erring on the side of including fewer galaxies in the sample.
In section \ref{data:imaging:sample} we discuss the cuts used for our source
sample.

\subsubsection{Photometric Redshifts} \label{data:imaging:photoz}

A photometric redshift (photoz) is estimated for each object in the source
catalog.  The repaired template fitting method is used, described in detail in
\cite{Csabai2000} and implemented in the SDSS EDR \citep{Csabai2003} as well as
the DR1 \citep{Abaz03}.  This technique uses the 5-band photometry for each
object as a crude spectrum.  The algorithm compares this spectrum to templates
for different galaxy types at different redshifts.  The result is an estimate
of the type and redshift of each galaxy.

There is a large covariance between the inferred type of the galaxy and its
photometric redshift.  The code outputs a full covariance matrix for type and
redshift.  Because we do not use the type information, we use the error
marginalized over type. We further assume that the resulting error is Gaussian,
which is only a good approximation for high S/N measurements.
This introduces a bias in the estimate of $\langle \Sigma_{crit}^{-1} \rangle$,
but Monte Carlo simulations indicate that this is a negligible effect.

From comparisons to galaxies with known redshifts, the {\it rms} in the SDSS
photoz estimates is found to be $\sim 0.035$ for \rmag\ $< 18$, increasing to
$\sim 0.1$ for \rmag\ $< 21$.  About 30\% of our source galaxy sample has
\rmag\ between 21 and 22.  Although the photozs are less reliable in this
magnitude range, these objects receive relatively little weight in the
analysis, because they have large shape errors and large photoz errors (and
hence small $\langle \Sigma^{-1}_{crit} \rangle$).

\begin{figure}[tbp]
\plotone{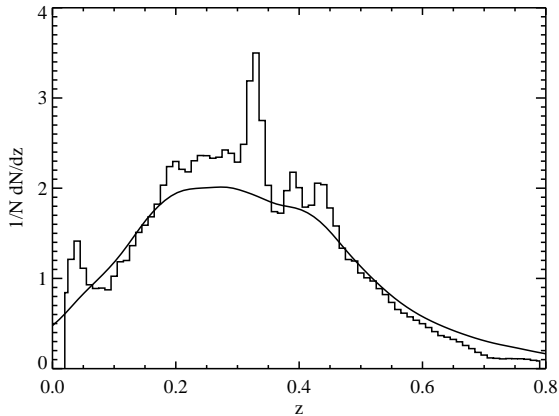} \figcaption{ Distribution of
photometric redshifts for sources used in this study.  The histogram shows the
photozs in bins of $\Delta z=.01$, and the smooth curve is derived from summing
the Gaussian distributions associated with each object.\label{fig:photozdist} }
\end{figure}

The photoz distribution for the sources used in this study is shown in figure
\ref{fig:photozdist}.  The histogram shows the photometric redshifts, and the
smooth curve is the distribution calculated by summing the Gaussian
distributions for each galaxy.  We use this smooth curve as a prior on the
photometric redshift when calculating the inverse critical density for each
lens-source pair (see \S \ref{lensing:deltasig}).  The large peaks in the
distribution are most likely not real, but rather the result of degeneracies in
the photometric redshift estimation.  This issue is addressed further in the
appendix.

Although there is significant overlap between the distribution of photozs and
the distribution of lens redshifts (figure \ref{fig:zhist}), sources with
photozs in front of or near the lens redshift are given appropriately
small weight according to equation \ref{eq:meansigcrit}.

\subsubsection{Defining the Source Sample} \label{data:imaging:sample}

\begin{figure}[hbp]
\plotone{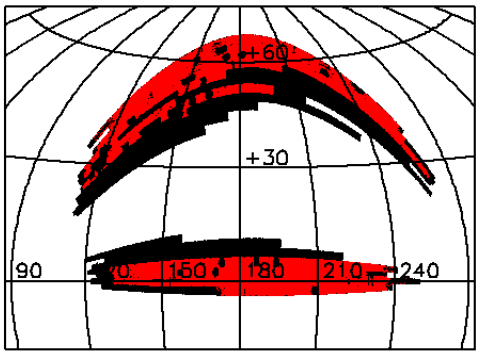} \figcaption{ Aitoff projection showing
the positions of the sources (black) and lenses (\clraitoff) used in this
study.  The section covering the equator is stripes 9-15.  The higher latitude
section is stripes 27-37.
\label{fig:aitoff}}
\end{figure}

Source galaxies are drawn from SDSS imaging stripes 9-15 and 27-37, covering a
region of nearly 3800 square degrees.  An Aitoff projection displaying the
positions of these sources (as well as the lenses) is shown in figure
\ref{fig:aitoff}. We make a series of cuts aimed at ensuring that the sample is
of high purity (free from stellar contamination) and includes only
well-resolved objects with usable shape information. We first require that the
extinction-corrected \rmag-band Petrosian magnitude is less than 22.  We next
make an object size cut, requiring that the resolution parameter $R>0.2$.  This
removes most of the stars and unresolved galaxies from the sample. However, at
faint magnitudes (\rmag$ > 21$), many stars and galaxies have similar values of
$R$ due to measurement error, so a further cut is needed.  We employ the
Bayesian galaxy probability (\S \ref{data:imaging:sg}) and find that the
combination $R > 0.2$, $P_g > 0.8$ guarantees that the source galaxy catalog is
greater than 99\% pure for \rmag\ $< 21.5$ and greater than 98\% pure for $21.5
<$\rmag$< 22$.

Additionally, we remove about $\sim$ 8\% of the sources---those with photoz
errors greater than 0.4, and we further exclude objects with photoz less than
0.02 or greater than 0.8 since failed measurements tend to pile up at a photoz
of 0.0 or 1.0.  This removes another 10\% of the objects.  The final source
catalog contains \numsource\ galaxies, corresponding to a density of about
$1-2$ source galaxies per square arcminute, depending on the local seeing.

\subsection{Spectroscopic Data} \label{data:spectro}

The lens galaxies are selected from the SDSS ``main'' galaxy spectroscopic
sample, which is magnitude- ($14.5 <$\rmag$< 17.77$) and surface
brightness-limited ($\mu_r < 23.5$), although these limits varied during the
commissioning phase of the survey.  See \citet{Strauss02} for a description of
``main'' galaxy target selection.

SDSS spectroscopy is carried out using 640 optical fibers positioned in
pre-drilled holes in a large metal plate in the focal plane.  Targeted imaging
regions are assigned spectroscopic plates by an adaptive tiling algorithm
\citep{Blanton03}, which also assigns each object a fiber.  The spectroscopic
data are reduced to 1-d spectra by \spectrotwo, and the \spectroone\ pipeline
outputs redshift and an associated confidence level, spectral classification
(galaxy, quasar, star), line measurements, and spectral type for galaxies,
among other parameters \citep{Stough02}.  In addition, the velocity dispersion
is measured for a large fraction of the early type galaxies.

For this analysis, we use a subset of the available spectroscopic ``main''
galaxy sample known as LSS \texttt{sample12} (M. Blanton 2003, private
communication). Although we draw from a larger sample, the mask (see \S
\ref{data:spectro:sphpolymasks}) was produced for this subset.  This sample is
also being used for analysis of the galaxy auto-correlation function
\citep{Zehavi04}, while a slightly earlier sample (\texttt{sample11}) has been
used to estimate the galaxy power spectrum \citep{Tegmark04}.  Using this
sample allows us to make meaningful comparisons between the auto-correlation
function and the galaxy-mass cross correlation function.  The spectroscopic
reductions used here are those of the \spectroone\ pipeline, for which
redshifts and spectroscopic classifications differ negligibly from those in the
above references.

\subsubsection{Redshifts} \label{data:spectro:z}

\begin{figure}[htbp]
\plotone{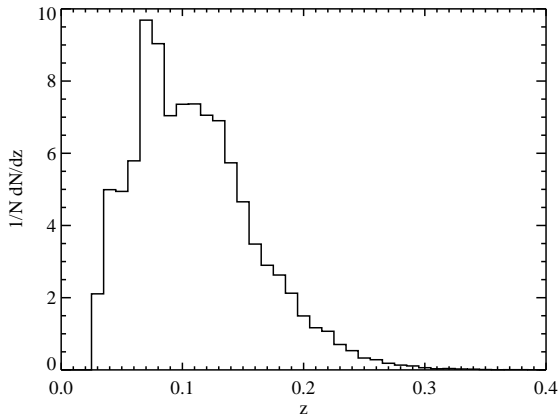} \figcaption{
Redshift distribution for galaxies used
in this study.  This sample contains only ``main'' galaxy
targets.\label{fig:zhist}}
\end{figure}

The SDSS spectra cover the wavelength range 3800-9200\AA\ with a resolving
power of 1800 \citep{Stough02}. Repeated 15-minute exposures (totaling at
least 45 minutes) are taken until the cumulative median (S/N)$^2$ per pixel in
a fiber aperture is greater than 15 at $g=20.2$ and $i=19.9$ in all 4
spectrograph cameras. Redshifts are extracted with a success rate greater than
99\%, and redshift confidence levels are greater than 98\% for 95\% of the
galaxies. Repeat exposures of a number of spectroscopic plates indicates that
``main'' galaxy redshifts are reproducible to 30 km/s.  We apply a cut on the
redshift confidence level at $>75$\%, which removes 0.5\% of the galaxies.
Many galaxies are further removed during the lensing analysis, as discussed in
\S \ref{data:spectro:sphpolymasks} and \S \ref{data:spectro:pixelmasks}.  A
redshift histogram is shown in figure \ref{fig:zhist} for the remaining
\numspec\ galaxies used in this work. The mean redshift for our sample using
the relative weights from the lensing analysis is $\langle z \rangle = 0.1$.

\subsubsection{K-corrections} \label{data:spectro:kcorr}

\begin{figure}[bp]
\plotone{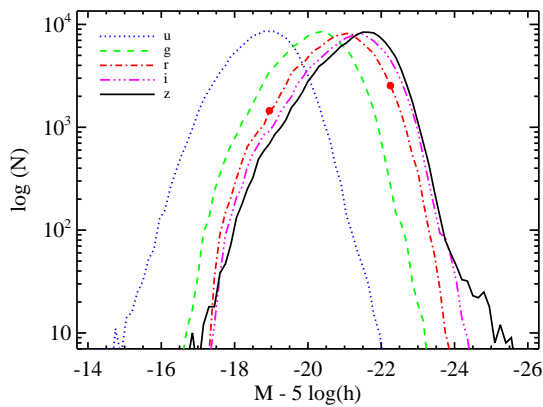} \figcaption{ Distribution of rest-frame absolute
magnitude in each of the 5 SDSS bandpasses for lenses used in this study. The
filled circles mark the magnitude range for the flux limited sample of
\citet{Zehavi03}.  Ninety percent of our lens sample falls within this range.
\label{fig:absmag_hist}}
\end{figure}

We apply K-corrections using the method discussed in \citet{Blanton02}
(\texttt{kcorrect v1\_10}).  Linear combinations of four spectral templates are
fit to the five SDSS magnitudes for each galaxy given its redshift.  Rest-frame
absolute magnitudes and colors are then calculated.  Figure
\ref{fig:absmag_hist} shows the distribution of absolute Petrosian magnitude
for each of the 5 SDSS bandpasses.

\subsubsection{Spectroscopic Masks} \label{data:spectro:sphpolymasks}

Because SDSS spectroscopy is taken through circular plates with a finite number
of fibers of finite angular size, the spectroscopic completeness varies across
the survey area. The resulting spectroscopic mask is represented by a
combination of disks and spherical polygons \citep{Tegmark04}.  Our spherical
polygon mask contains 3844 polygons covering an area of 2818 square degrees.
Each polygon also contains the completeness, a number between 0 and 1 based on
the fraction of targeted galaxies in that region which were observed.  We apply
this mask to the spectroscopy and include only galaxies from regions where the
completeness is greater than 90\%.  The same criteria are used to generate the
random points as discussed in \ref{gglmeas:random}.

\subsubsection{Photometric Masks} \label{data:spectro:pixelmasks}

Although the correction for PSF smearing (\S \ref{data:imaging:correct})
substantially reduces the galaxy shape bias, it does not completely eliminate
it---generally, a small, slowly varying residual remains.  Fortunately, a
residual PSF bias that is constant over the lens aperture cancels on average
from the azimuthally averaged tangential shear, since two aligned sources
separated by 90\degr\ relative to the lens contribute equally but with opposite
sign. To take advantage of this cancellation, we divide the source galaxies
around each lens into quadrants and demand that at least two adjacent quadrants
are free of edges and holes out to the maximum search radius.

We represent the geometry of the sources using the hierarchical pixel scheme
SDSSPix\footnote[1]{http://lahmu.phyast.pitt.edu/$\sim$scranton/SDSSPix}
\citep{Scranton03}, modeled after a similar scheme developed for CMB analysis
\citep{Gorski98}. This scheme represents well the rectangular geometry of the
SDSS stripes.  The mask is a collection of pixels at varying resolution
covering regions with holes or edges.  We do not mask out bright stars (which
cover only a tiny fraction of the survey area) for the lensing analysis.

After checking the spherical polygon masks (\S
\ref{data:spectro:sphpolymasks}), each lens galaxy is checked against the pixel
mask to guarantee that it is within the allowed region.  Each quadrant around
the lens is then checked to determine if it contains a hole or an edge.  Lenses
are excluded from the sample if there are no adjacent quadrants that are
completely unmasked.  In addition to this cut, we demand that the angular
distribution of source galaxies around the lens have ellipticity no greater
than 20\% in order to ensure the availability of pairs with 90\degr\
separation.  The same criteria are applied to the random points (see \S
\ref{gglmeas:random}).

We draw the final spectroscopic data set from SDSS stripes $9-12$ (near the
equator) and $28-37$.  We do not use galaxies from the 3 southern stripes
(76,82,86), because they contain few lens-source pairs at large separation.
Figure \ref{fig:aitoff} shows the distribution of the lens and source galaxies
as an Aitoff projection.  After applying the cuts described above, the final
lens sample contains \numspec\ galaxies.

\section{Results: The Mean \deltasig\ from $.02-10\MakeLowercase{h}^{-1}$ M\MakeLowercase{pc}} \label{gglmeas:deltasig}

\begin{figure}[htbp]
\plotone{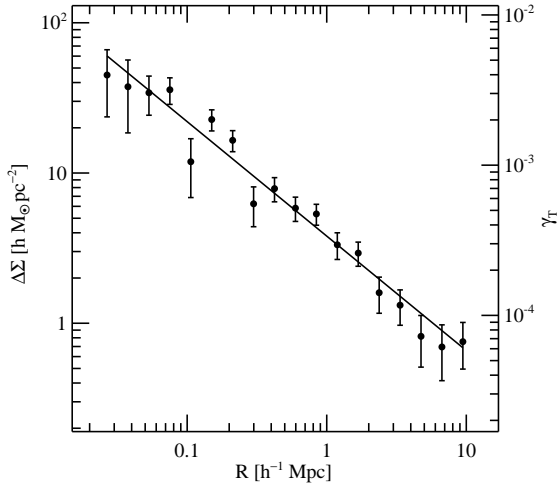} \figcaption{ Mean \deltasig\ = $\overline{\Sigma}(<R) -
\overline{\Sigma}(R)$ measured for the full lens sample. The solid line is the
best-fitting power law \deltasig$\varpropto R^{-0.76}$.  The right axis shows
the corresponding tangential shear $\gamma_T$.
\label{fig:deltasig} }
\end{figure}

The mean lensing signal \deltasig\ for the full sample of SDSS lens galaxies is
shown in figure \ref{fig:deltasig}.  The signal is unambiguously detected from
25 $h^{-1}$ kpc to 10 $h^{-1}$Mpc.  The corresponding mean shear $\gamma_T$ is
shown on the right axis. Corrections have been made to this profile as
described in \S \ref{gglmeas:random}. These corrections are relatively small in
all radial bins.

\begin{figure}[bp]
\plotone{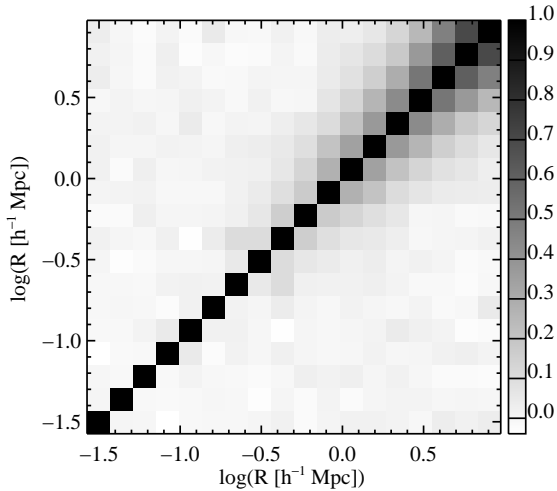} \figcaption{
Correlation matrix for \deltasig\ in figure \ref{fig:deltasig} calculated
using jackknife re-sampling. 
\label{fig:corr_matrix_deltasig} }
\end{figure}

The errors in figure \ref{fig:deltasig} come from jackknife re-sampling.
Although we expect statistical errors to dominate over sample variance even on
the largest scales shown here, there are in addition large variations in the
systematic errors.  Due to gaps between the 5 CCD columns, two interleaving
imaging runs make up a contiguous imaging stripe, and they are generally taken
on different nights under different photometric conditions. As a result, the
residuals from the PSF correction vary between the columns of interleaving
runs, each of which is $\sim 0.2$ degrees wide.  The residuals also vary over
time along the direction of the scan, with a typical scale of a few degrees.
Thus the proper subsample size to account for this variation is about a square
degree.  We divide the sample into 2,000 disjoint subsamples, each
approximately a square degree in size (see \S \ref{data:spectro:sphpolymasks})
and remeasure \deltasig\ 2,000 times, leaving out each subsample in
turn. Figure \ref{fig:corr_matrix_deltasig} shows an image of the resulting
dimensionless correlation matrix $Corr_{i,j} = V_{i,j}/\sqrt{V_{i,i}V_{j,j}}$,
where $V_{i,j} = \langle (\Delta \Sigma(R_i) - \langle \Delta
\Sigma(R_i)\rangle)(\Delta \Sigma(R_j)-\langle \Delta \Sigma(R_j)
\rangle)\rangle$.  The off-diagonal terms are negligible in the inner bins but
become important beyond $R \sim 1 h^{-1}$ Mpc.  The full covariance matrix is
used for all model fitting.

The mean \deltasig\ for the full sample is well described by a power law 
\begin{equation}
\Delta\Sigma(R) = A (R/1\textrm{Mpc})^{-\alpha}
\end{equation}
with $\alpha = 0.76 \pm 0.05$ and $A = (3.8 \pm 0.4) h
$M$_{\sun}$pc$^{-2}$. The outliers at intermediate radii make the reduced
$\chi^2$ for the power law fit somewhat poor but not unacceptable: $\chi^2/\nu
= 1.26$.  There is a 20\% chance of $\chi^2$ exceeding this value
randomly.

\begin{figure}[bp]
\plotone{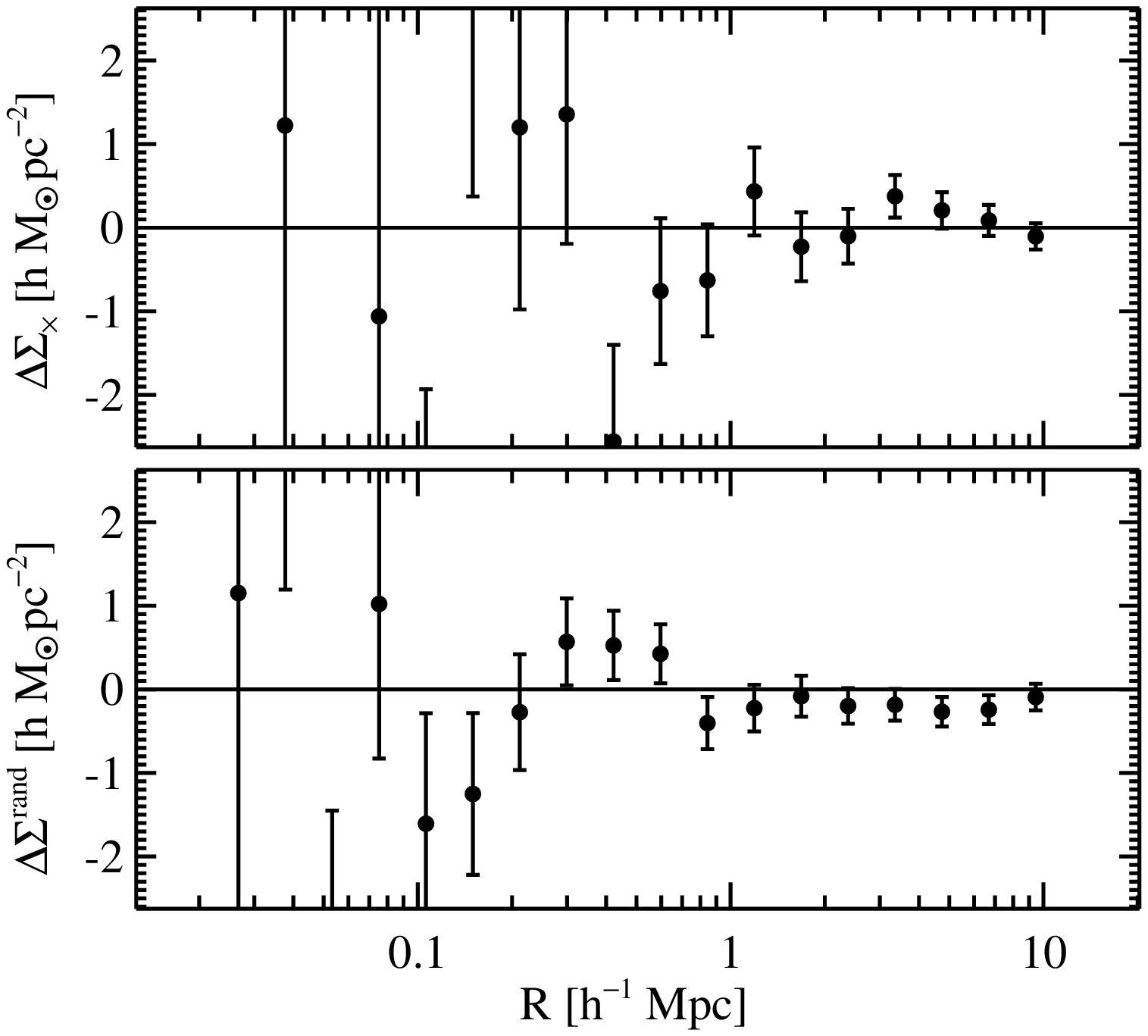} \figcaption{ Two tests for systematics in the
lensing measurement. The top panel shows the ``B-mode'' for lensing,
\deltacross, measured around the same lenses used in figure \ref{fig:deltasig}.
The measurement is consistent with zero, as expected for lensing. The bottom
panel shows \deltasig\ measured around \numrand\ random points.  The detection
at large radius is indicative of systematic errors, most likely from residuals
in the PSF correction.  This has been subtracted from the signal around lenses
for figure \ref{fig:deltasig}.
\label{fig:rand_and_ortho} }
\end{figure}

An important check for systematic errors is the ``B-mode'', \deltacross\, the
average shear signal measured at 45\degr\ with respect to the tangential
component.  If the tangential shear signal is due solely to lensing, the B-mode
should be zero, whereas systematic errors generally contribute to both the E-
and B-modes.  The top panel of figure \ref{fig:rand_and_ortho} shows
\deltacross\ measured using the same source-lens sample used in figure
\ref{fig:deltasig}. This measurement is consistent with zero.

\subsection{Systematics Tests with Random Points} \label{gglmeas:random}

By replacing the lens galaxies with sets of random points, we can gauge two
systematic effects on \deltasig: residuals in the PSF correction and the radial
bias due to clustering of sources with the lenses. We generated a random sample
with ten times as many points as the lens sample, using the same masks and
selection criteria described in \S \ref{data:spectro:sphpolymasks} and \S
\ref{data:spectro:pixelmasks}. The random points are drawn from the same
redshift distribution as the lens galaxies.  These criteria guarantee that the
same regions, and thus roughly the same systematics, are sampled by the lenses
and the random points.  Any non-zero lensing signal for the random points we
ascribe to residuals in the PSF correction.

The bottom panel of figure \ref{fig:rand_and_ortho} shows \deltasig\ measured
around \numrand\ random points, with errors from jackknife re-sampling.  Note
that the random sample is large enough that in this case sample variance
dominates the error in the outer bins. There is a significant signal at large
radius.  The signal at smaller radii is less well determined, but it is in any
case far below the signal due to lenses.  Interpreting the large-scale signal
as systematic error, we subtract it from the \deltasig\ measured around lenses
and add the errors $\sigma_{final}^2 = \sigma_{lens}^2 + \sigma_{rand}^2$; this
correction is incorporated in figure \ref{fig:deltasig}.

\begin{figure}[bp]
\plotone{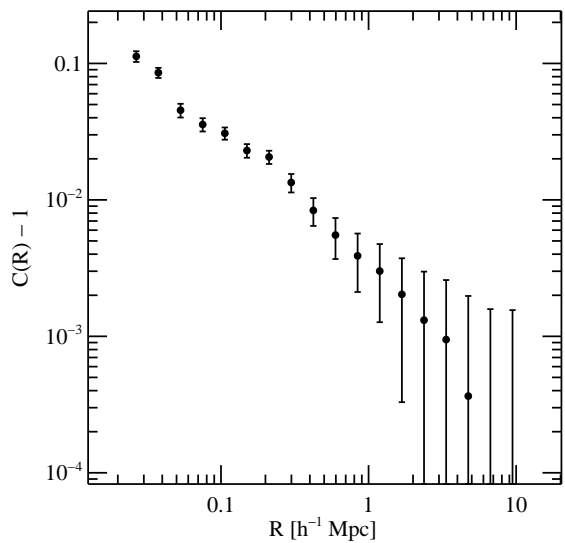} \figcaption{ Correction factor for the clustering of
sources around the lens galaxies.  The function $C(R)-1$ is essentially a
weighted cross-correlation function between lenses and sources.  The
\deltasig\ in figure \ref{fig:deltasig} has been multiplied by $C(R)$, which
is a negligible correction for radii larger than $\sim$ 50 kpc.
\label{fig:cluster_corr} }
\end{figure}

The second systematic probed by the random sample involves clustering of the
source galaxies with the lenses. The calculation of the mean inverse critical
density in \S \ref{lensing:deltasig} properly corrects for the fact that a
fraction of the source galaxies are in front of the lenses, but only under the
assumption that the lens and source galaxies are homogeneously
distributed. Since galaxies are clustered, a small fraction of the sources are
in fact physically associated with the lenses, causing a scale-dependent bias
of the lensing signal.  We correct for this by estimating the excess of sources
around lenses compared with the random points.  The correction factor is the
ratio of the sums of the weights for sources around lenses and around random
points:
\begin{equation} \label{eq:cluster_corr}
C(R) = \frac{N_{rand}}{N_{lens}} 
\frac{\sum_{i,j} w_{i,j}}{ \sum_{k,l} w_{k,l}}
\end{equation}
where $i,j$ indicates sources found around lenses, $k,l$ indicates sources
found around random points, and $w_{i,j}$ is the weight for the lens-source
pair (see \S \ref{lensing:deltasig}). The value of $C(R)-1$ is shown for the
full sample in figure \ref{fig:cluster_corr}.  The use of photometric redshifts
reduces the correction significantly, since sources associated with the lenses
are given little weight.  We find that $C(R)-1$ falls roughly as
$R^{-0.9}$. The correction for clustered sources is essentially negligible
beyond 50 kpc.  The signal in figure \ref{fig:deltasig} has been multiplied by
$C(R)$.

We measure this correction factor separately for each of the lens subsamples
presented in later sections.  Since galaxy clustering increases with
luminosity, and higher luminosity lens galaxies are seen out to higher redshift
where more of the faint sources are near the lenses, the correction factor
$C(R)$ increases with the luminosity of the lenses.  The correction for our
highest luminosity samples is 2.0 at 25 kpc compared to 1.1 for the full
sample. Similarly, the correction for early type lens galaxies is larger,
while the correction for late types is smaller than for the full sample.
Although the corrections for some samples are large, the value of the
correction is well measured in each case.

We have also calculated the average \deltasig\ for the luminosity subsamples
defined below (\S \ref{gmcflum}).  Because $C(R)$ depends on luminosity, the
average of the subsamples could in principle differ from the mean \deltasig\
estimated from the full sample, as suggested by \citet{Guzik02} regarding the
results of \citet{Mckay02}.  However, we find no significant difference in the
two methods for our weighting scheme, which results in smaller corrections than
those in \citet{Mckay02}.

Note our calculation of $C(R)$ may be a slight over- or under-estimate, since
lensing itself may induce some correlation between unassociated background
sources and the positions of the foreground lenses through magnification.
Lensing magnification can bring galaxies into a magnitude limited sample that
would otherwise have been too faint to be included, and this effect will be a
function of scale.  On the other hand, the geometric distortion induced by the
lens moves the apparent position of background galaxies radially outward,
decreasing the number density.  The net change in number density depends on
magnification $\mu$ and the slope of the galaxy number counts $s$: the ratio of
counts with and with out lensing is $N/N_0 \varpropto \mu^{(2.5 s - 1)}$
\citep{Broad95}.  The magnification is of the same order as the shear, which
for our lens sample is $\lesssim 10^{-3}$, and the slope $s$ is typically about
0.4, which results in negligible magnification bias.

\section{The Galaxy-mass Correlation Function} \label{gglmeas:xigm}

\subsection{Power-law Fits to \deltasig}

As noted in \S \ref{lensing:shear}, a power-law \deltasig\ is consistent,
within the errors, with the galaxy-mass correlation function $\xi_{gm}$ also
being a power-law, $\xi_{gm} = (r/r_0)^{-\gamma}$, with slope $\gamma =
1+\alpha$.  Fitting for $\gamma$ and $r_0$, we find best fit marginalized
values of $\gamma = 1.76 \pm 0.05$ and $r_0 = (5.7 \pm
0.7)(0.27/\Omega_m)^{1/\gamma} h^{-1}$ Mpc. Because \deltasig\ is proportional
to $\overline{\rho}$, we have assumed a fiducial value for the mean density
$\overline{\rho}=\rho_{crit} \Omega_m$ determined by WMAP + ACBAR + CBI in
combination with the power spectrum from 2dFGRS and Lyman $\alpha$ data, which
yields $\Omega_m = 0.27 \pm 0.02$ \citep{Spergel03}.  Marginalizing over this
small uncertainty in $\Omega_m$ does not change our error estimates.

\subsection{Inversion to \xigm}

\begin{figure}[htbp]
\plotone{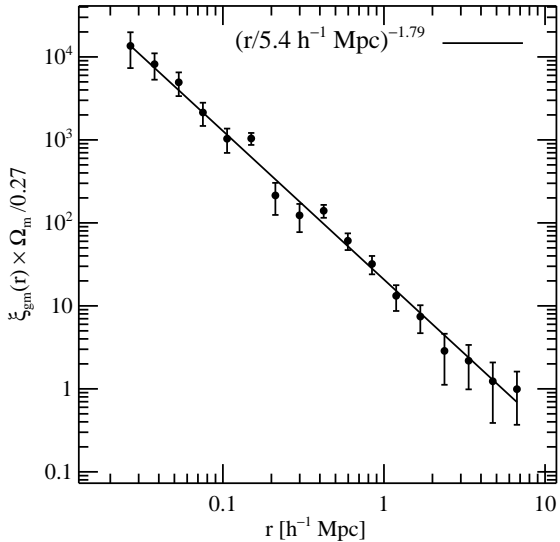} \figcaption{Galaxy-mass cross-correlation function
\xigm, obtained by inverting \deltasig.  The amplitude of \xigm\ is degenerate
with $\Omega_m$, which we have set to 0.27.  The solid line is the best-fit
power law $\xi_{gm} = (r/5.4 h^{-1}$ Mpc$)^{-1.79}$.
\label{fig:xigm}
}
\end{figure}

In addition to fitting a power law to \deltasig, we performed a direct
inversion of \deltasig\ to \xigm\ as outlined in \S \ref{lensing:shear}.
\xigm\ for the main sample is shown in Figure \ref{fig:xigm}.  Again, since our
measurements scale linearly with $\Omega_m$, we have assumed a fiducial value
of $\Omega_m = 0.27$.

In principle, the inversion from \deltasig\ to \xigm\ at a given radius $r$
requires data for \deltasig\ to $R=\infty$.  In practice, the contribution from
scales beyond our data is negligible in all radial bins but the last few.  We
estimate the contribution from large scales by extrapolating the best-fit power
law.  Because the last point would be based entirely on the extrapolation, we
exclude it from figure \ref{fig:xigm}.  The contribution of the extrapolation
is small for the remaining points except for the last bin at 6.7 Mpc, for which
it is a 40\% effect.

Because we do not have any a priori knowledge of \xigm\ at large scales, we
must include this extrapolation in the error budget.  A lower bound on \xigm\
is obtained by assuming there is no signal beyond $R_{max}$ at all:
$\xi_{gm}^{low} = \xi_{gm}^{interior}-\sigma(\xi_{gm})$, where
$\sigma(\xi_{gm})$ is the error on $\xi_{gm}^{interior}$ propagated from
\deltasig.  The most conservative error estimate is then $|\xi_{gm} -
\xi_{gm}^{low}|$. This is the error plotted in figure \ref{fig:xigm} and is
used for scaling the correlation matrix discussed below.  Again, this only
makes an significant difference in the last radial bin.  This may still
underestimate the upper bound on \xigm\ if in fact $\xi_{gm}$ is considerably
flatter at separations larger than $R_{max}$.  

Due to the finite binning of \deltasig\ (which is well described by a power
law), linear interpolation results in a slight overestimate of the derivative
$d \Delta\Sigma/dR$.  A small $(\sim 4\%)$ correction for this effect has been
applied in figure \ref{fig:xigm}.

\begin{figure}[bp]
\plotone{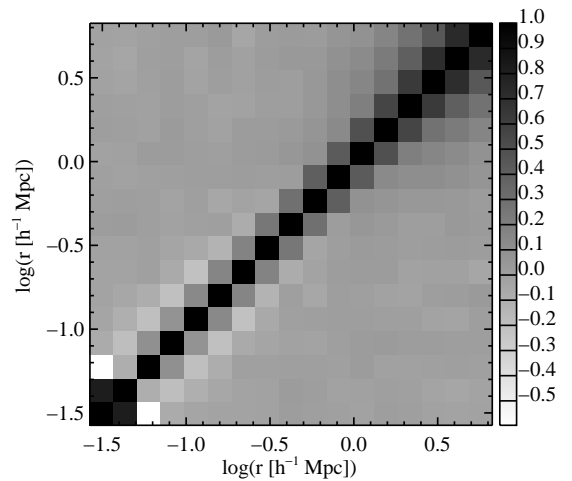} \figcaption{
Correlation matrix for \xigm\ in figure \ref{fig:xigm}, propagated from the
covariance matrix of \deltasig. 
\label{fig:corr_matrix_xi} }
\end{figure}

Because the inversion is a linear operation on \deltasig, it is straightforward
to obtain the covariance matrix for \xigm.  The corresponding dimensionless
correlation matrix is shown in figure \ref{fig:corr_matrix_xi}.  Due to the
differentiation and integration involved in the inversion, adjacent points of
\xigm\ are strongly correlated; i.e., the resulting covariance matrix for
\xigm\ is less diagonal than the covariance matrix for \deltasig.  The full
covariance matrix is used for the model fits discussed below.

We fit \xigm\ to a power law model $\xi_{gm} = (r/r_0)^{-\gamma}$. For the
inversion to \xigm\ we cannot use the last radial bin, so this fit only applies
for $R < 6.7$ Mpc.  The results are summarized in table \ref{tab:allsamp}. We
find best-fit parameters $r_0 = (5.4 \pm 0.7)(0.27/\Omega_m)^{1/\gamma}$ and
$\gamma = 1.79 \pm 0.06$, which are in agreement with the results we obtained
fitting directly to \deltasig.  These results are also in excellent agreement
with the auto-correlation function from \citet{Zehavi03}.  In the next section,
we compare \xigm\ and \xigg\ directly to constrain the bias $b/r$.

\section{Bias} \label{gglmeas:bias}

The ratio of the auto-correlation function \xigg\ to the galaxy-mass
correlation function \xigm\ is a measure of the bias between galaxies and mass
$\xi_{gg}/\xi_{gm} = b/r$ (see \S \ref{intro}).  It is important to use the
same galaxies for both \xigm\ and \xigg.  We use the auto-correlation function
from \citet{Zehavi03} (hereafter Z03), measured for a flux limited sample of
118,000 SDSS ``main'' galaxies. The selection function for the Z03 sample is
very similar to that of our lens sample, but the former covers a more
restricted range of absolute magnitudes: $-22.2<~M_r~<-18.9$, while our sample
spans the range $-24 <~M_r~< -17$. However, figure \ref{fig:absmag_hist} shows
that 90\% of the galaxies in our lens sample in fact lie within the Z03
magnitude range.  Also, while each galaxy is weighted differently in the two
correlation function measurements, the mean luminosities of the two samples are
quite similar: for the Z03 sample, $\langle L_r \rangle = 1.7\times 10^{10}
h^{-2} L_{\sun}$, while the mean for our sample is $\langle L_r \rangle
=1.46\times 10^{10} h^{-2} L_{\sun}$, both comparable to $L_* = 1.54\times
10^{10} h^{-2} L_{\sun}$ \citep{blanton01}.  Assuming that $r_0$ for $\xi_{gg}$
scales with luminosity as in \citet{Norberg01}, we expect the auto-correlation
length for our sample of galaxies to be only slightly lower than that of Z03,
$r_0 = 5.60$ rather than $5.77 h^{-1}$ Mpc.

\begin{figure}[t]
\plotone{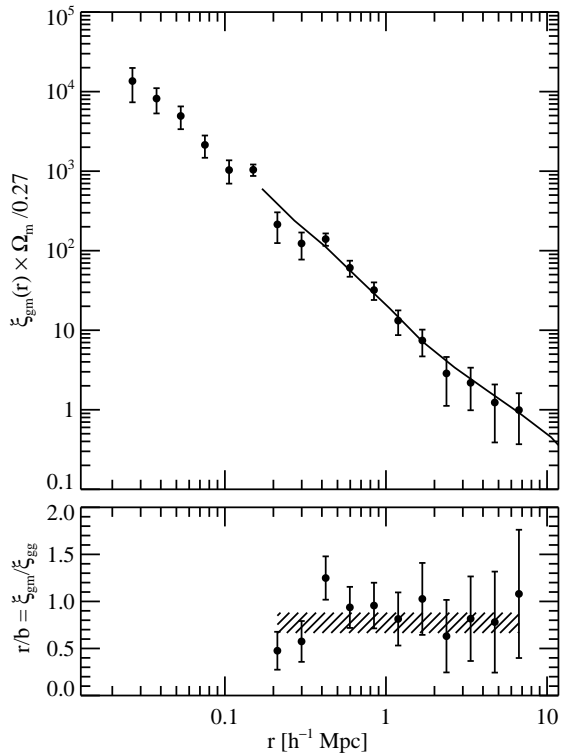} \figcaption{Galaxy-mass correlation function
(filled circles) and the galaxy auto-correlation function (solid line) for
similarly selected samples of SDSS galaxies.  The top panel shows the two
correlation functions and the bottom panel shows the inverse galaxy-mass bias
$r/b = \xi_{gm}/\xi_{gg}$. We plot the inverse because \xigm\ is much
noisier than \xigg. The bias is consistent with scale independent, with a mean
of $\langle b/r \rangle = (1.3 \pm 0.2)(\Omega_m/0.27)$, shown as the
hashed region.
\label{fig:xigm_xigg} }
\end{figure}

Figure \ref{fig:xigm_xigg} shows \xigm\ (points with error bars) and the \xigg\
from Z03 (solid line).  The \xigg\ was inferred by inverting from \wgg\ using
the same techniques outlined in \S \ref{lensing:shear} (I. Zehavi, 2003,
private communication).  The amplitude of \xigg\ has been lowered by about 5\%
to account for the slight difference in mean luminosity for the two samples as
discussed above.  $\xi_{gg}$ and $\xi_{gm}$ for these samples are 
consistent within the errors over the common range of radii.  The inverse
galaxy-mass bias $r/b = \xi_{gm}/\xi_{gg}$ is shown in the bottom panel
of figure \ref{fig:xigm_xigg}. We use the inverse because $\xi_{gm}$ is much
noisier than $\xi_{gg}$. To calculate $b/r$, we have interpolated the points
and errors from Z03. The bias $b/r$ is consistent with unity and is scale
independent within the errors.  Fitting the bias to a constant over the radial
range $0.2-6.7 h^{-1}$ Mpc, we find
\begin{equation}
\langle b/r \rangle = (1.3 \pm 0.2)\left(\frac{\Omega_m}{0.27}\right).
\end{equation}

\section{Galaxy-mass Correlations in a Volume Limited Sample of Luminous Galaxies} \label{gglmeas:vlim}

In the previous section we used a flux limited sample to measure \xigm.  This
selection has the advantage that it uses the largest possible sample of lens
galaxies and thus provides the best possible S/N measurement of \xigm.  The
interpretation is somewhat complicated, however, by the selection function. In
order to provide a measurement that is simpler to interpret, we also defined a
volume and magnitude limited sample, using the limits for the high-luminosity
sample of \citet{Zehavi02}: $0.1 < z < 0.174$, $-23.0 < M_r -5 \log(h) <
-21.5$, $\langle L_r \rangle = 3.96 \times 10^{10} L_{\sun}$. This sample
contains \numspecvlim\ lenses.

\begin{figure}[b]
\plotone{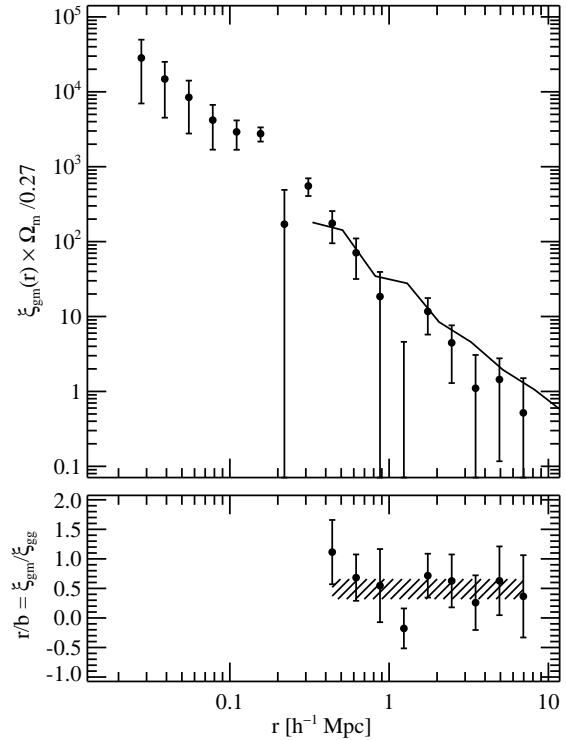} \figcaption{
Same as figure \ref{fig:xigm_xigg}, but for a volume and magnitude limited
sample $(0.1 < z < 0.174)$, $(-23.0 < M_r -5 \log(h) < -21.5)$.  The solid line
is \xigg\ for the same selection criteria from \citet{Zehavi02}.  We detect no
scale dependence in the bias $b/r$.  The mean bias over the common range of
scales ($0.4-6.7 h^{-1}$ Mpc) is $\langle b/r \rangle_{Vlim} = (2.0 \pm
0.7)(\Omega_m/0.27)$. 
\label{fig:vlim3_xigm_xigg}}
\end{figure}

The inferred \xigm\ for this sample is shown in the top panel of figure
\ref{fig:vlim3_xigm_xigg}, along with \xigg\ inferred from the same volume and
magnitude limited sample from \citet{Zehavi02}.  Power law fits to \xigm\ are
shown in table \ref{tab:allsamp}.  The slope is significantly steeper than for
the overall sample.  \xigm\ shows deviations from a pure power-law, however,
especially at small separations ($r < 0.2 h^{-1}$ Mpc), and this is reflected
in the poor reduced $\chi^2$ for the power law fit.  We will study the shape of
\xigm\ as a function of luminosity in more detail in \S \ref{gmcflum}, where we
split the overall sample into multiple bins in luminosity,

The bottom panel of figure \ref{fig:vlim3_xigm_xigg} shows the inverse bias
$r/b$ over the common range of scales: $0.4 < r < 6.7 h^{-1}$ Mpc.  The $b/r$
for this volume limited sample is consistent with scale-independent. The mean
bias over these scales is
\begin{equation}
\langle b/r \rangle_{Vlim} = (2.0 \pm 0.7)\left(\frac{\Omega_m}{0.27}\right)
\end{equation}
Because this is a higher luminosity sample than the overall flux limited
sample, one expects a larger bias \citep{Norberg01}, but we do not yet detect
any significant increase in the bias, as $\langle b/r \rangle_{Vlim}$ is
consistent with $\langle b/r \rangle$ at the 1-$\sigma$ level.

\section{Dependence of Galaxy-mass Correlations on Luminosity}
\label{gmcflum}

To study galaxy-mass clustering as a function of galaxy luminosity, we split
the lens sample into three bins of luminosity separately in each of the five
SDSS bandpasses, as shown in table \ref{tab:lumbin}. The binning is chosen
primarily to yield comparable lensing S/N in each bin. To achieve this we place
85\% of the galaxies in the lowest luminosity bin, 10\% in the middle, and 5\%
in the highest bin.  The low-luminosity bin must be so large because
intrinsically faint galaxies are weakly clustered and are only detected in the
SDSS at low redshift (for which the lensing efficiency is low). Since it
contains the majority of lens galaxies, the faintest subsample has a mean
luminosity comparable to the overall sample ($\sim L_*$), while the most
luminous subsample is $\sim 5-6$ times brighter. Note that the number in each
bin varies as a function of bandpass because galaxies are not detected in every
bandpass. Also note that, while the different bins in luminosity are
independent, the subsamples in the different bandpasses are simply re-samplings
of the same lenses.

\begin{figure*}[p]
\plotone{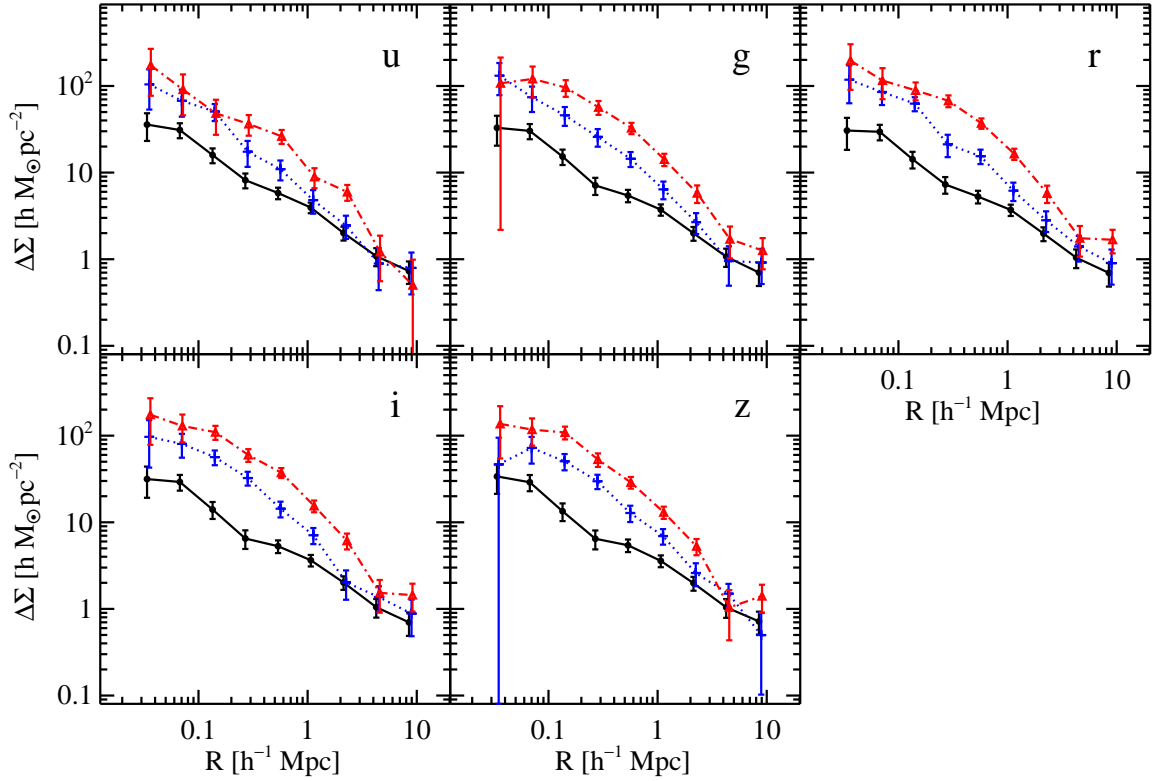} \figcaption{ Mean \deltasig\ in
three luminosity subsamples for each of the 5 SDSS bandpasses.  In each panel,
circles connected by solid lines\lumblack, crosses connected by dotted
lines\lumblue, and triangles connected by dot-dashed lines\lumred\ represent
measurements for the lowest, middle, and highest luminosity subsamples. The
data have been re-binned to 9 bins from 18 for clarity.
\label{fig:deltasig_allband_bylum} }
\end{figure*}

The mean \deltasig\ for each luminosity subsample is shown in Figure
\ref{fig:deltasig_allband_bylum}. The profiles have been re-binned from 18 to 9
radial bins for clarity of presentation. Since the lowest luminosity subsample
contains the large majority of lenses, the signal is comparable to that of the
overall sample in both amplitude and shape. The higher luminosity subsamples
show a significantly steeper slope on intermediate to large scales, and a
flattening at smaller scales.  This trend is seen in each of the bandpasses.

\begin{figure*}[p]
\plotone{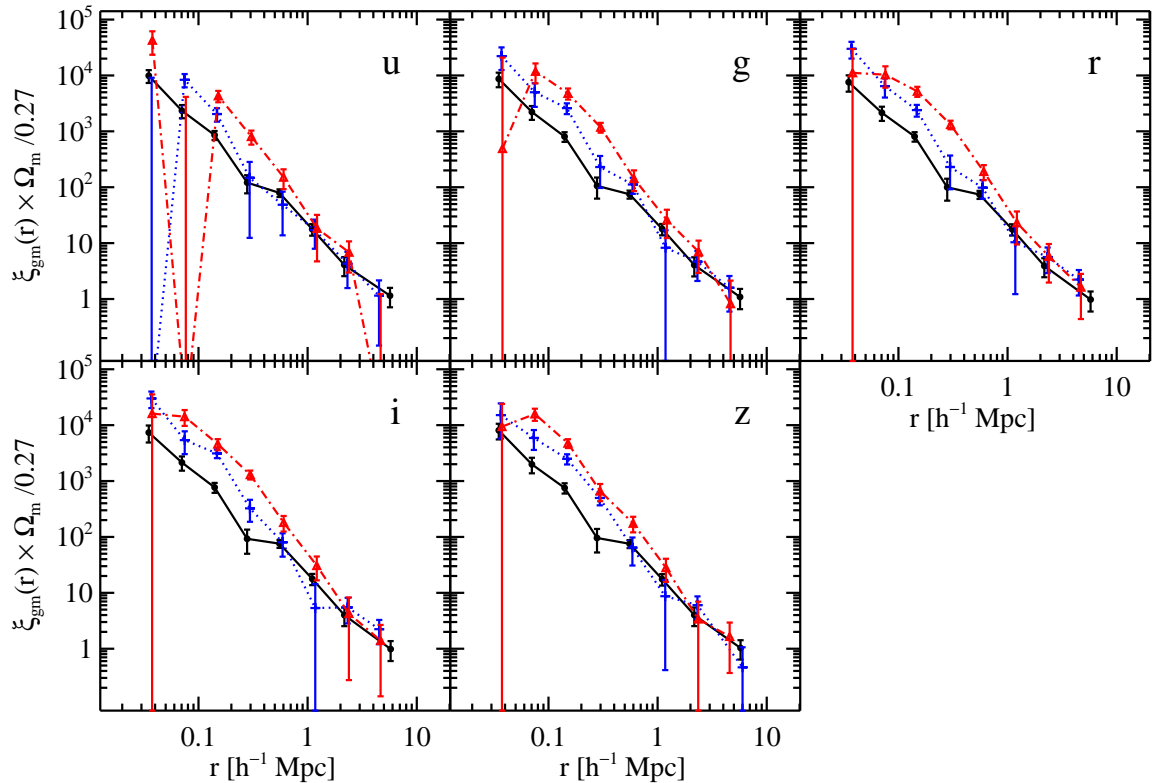} \figcaption{ Same as figure
\ref{fig:deltasig_allband_bylum}, but now plotting \xigm\ obtained by inverting
\deltasig.  The highest luminosity bin shows significant deviation from a power
law.
\label{fig:xi_allband_bylum} }
\end{figure*}

The \deltasig\ for each luminosity bin was inverted to \xigm\ in the same way
as for the overall sample.  The resulting \xigm\ for each subsample is shown in
figure \ref{fig:xi_allband_bylum}.  Note that the shape and relative amplitude
of these data are independent of the assumed $\Omega_m$.

\xigm\ for the highest luminosity bins deviate significantly from power laws,
although this trend is less clear than for \deltasig.
\xigm\ at large radius ($r > 1 h^{-1}$ Mpc) is consistent across the different
luminosity bins, while at intermediate radii ($0.1 < r < 1 h^{-1}$ Mpc) the
amplitude of the correlations increases with luminosity.  On the smallest
scales ($r < 0.1 h^{-1}$ Mpc), the clustering increases weakly with luminosity.
These trends are in qualitative agreement with the predictions from
semi-analytic and n-body results from \citet{Guzik01}. Note, the low data
points in \xigm\ for the middle (and highest) \umag-band luminosity bin is due
to a single negative point in the unbinned \deltasig.

We fit the \xigm\ to power law models $\xi_{gm} = (r/r_0)^{-\gamma}$ for each
of the luminosity subsamples.  The results for these fits, summarized in table
\ref{tab:lumbin}, indicate that \xigm\ for the low luminosity samples is a
good fit to a power law and consistent with the overall sample.  The two
highest luminosity bins have a steeper logarithmic slope than the low
luminosity sample.  The highest luminosity bin, however, is a poor fit to a
power law in each of the bandpasses.  The fits over all radius give a slope of
$\gamma \sim 2$.  The slope for $r > 0.1$ Mpc is $\gamma \sim 2.5$, much
steeper than the slope including the points at small radii. This is indicative
of the rollover seen at small separations.


A priori, one might hypothesize that the change in slope from $L_*$ to brighter
galaxies is partly due to the fact that the higher luminosity subsamples are
also redder: for the \rmag-band samples, $g-r = 0.73(0.77)$ for the
middle(brightest) sample, while $g-r = 0.63$ for the full sample (see table
\ref{tab:lumbin}).  However, the same increase of slope is also seen for the
\umag\ subsamples, for which the trend of $g-r$ color with luminosity is much
less pronounced.

\section{Dependence of Galaxy-mass Correlations on Galaxy Spectral Type and Color}
\label{gmcftype}


In addition to luminosity, we can also study how the lensing signal varies with
galaxy spectral type and color.  The \spectroone\ pipeline classifies galaxies
as early to late spectral types using the method developed in
\citet{Connolly95} and \citet{Connolly99}.  A Karhunen-Lo\`{e}ve decomposition
(KL) is performed on a large set ($\sim 100,000$) of galaxy spectra, ranging
from quiescent E-Sc types to starburst irregulars.  The spectrum of each galaxy
is expanded in terms of the KL eigenspectra, and the first five coefficients of
the expansion \{\texttt{ECOEFF1,..,ECOEFF5}\} are measured.  Since most of the
information is contained in the first two coefficients, a simple one-parameter
family, constructed from the angle in the plane of the first two eigenvectors,
provides a sufficient classification: \eclass\ =
$\tan^{-1}(-\texttt{ECOEFF2/ECOEFF1})$.  Early types have more negative
\eclass.

\begin{figure}[tbp]
\plotone{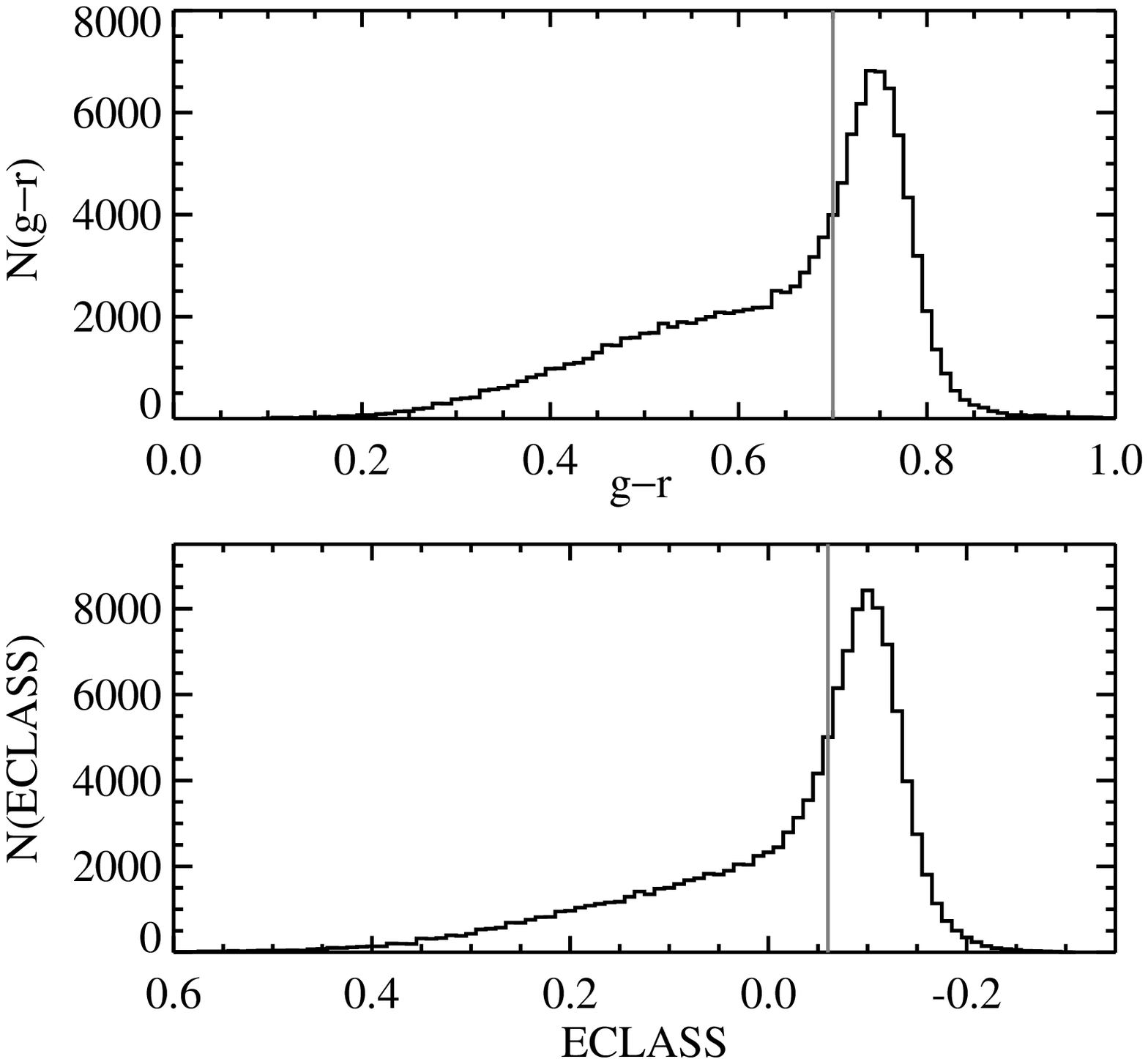} \figcaption{ Distribution of \gmr\ (top panel)
and \eclass\ (bottom panel) for lens galaxies.  
Early types have more negative \eclass, which has
been plotted in reverse for comparison with \gmr. We divide the
samples at \gmr\ = \gmrcut\ and \eclass=\eclasscut\ as shown by the gray
vertical lines. \label{fig:eclass}}
\end{figure}

The distribution of \eclass\ for the lens galaxy sample is shown in the bottom
panel of figure \ref{fig:eclass}.  The rest frame \gmr\ distribution is also
shown in the top panel.  The \gmr\ color and \eclass\ correlate well.  We make
cuts at \eclass=\eclasscut\ and \gmr\ = 0.7 to split the sample into early/late
types and red/blue.  These cuts each divide the full sample roughly in half.

\begin{figure}[t]
\plotone{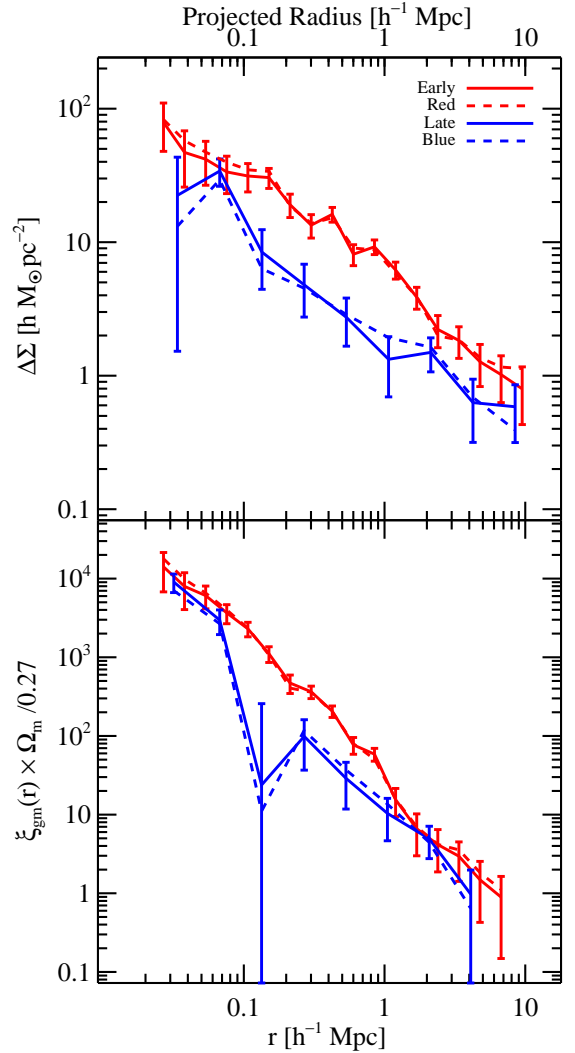} \figcaption{ Mean \deltasig\ and
\xigm\ for different classes of galaxies. The top panel shows \deltasig\ for
galaxies classified as early or red in thick solid and dashed lines
respectively\lumred.  Late type and blue galaxies are shown in thin solid and
dashed lines\lumblue. The trends for the \eclass\ and \gmr\ cuts are very
similar, as expected since color and spectral type are highly correlated.  The
bottom panel shows the inversion to \xigm.
\label{fig:deltasig_early_late} }
\end{figure}

The mean \deltasig\ for these different classes is shown in the top panel of
figure \ref{fig:deltasig_early_late}.  The red galaxies show a power law
$\Delta \Sigma$ for $R > 100$ kpc and a flattening at smaller separations,
similar to the two highest luminosity bins of the full sample.  The blue and
late type measurements, although noisy, are consistent with a power law
\deltasig\ over all separations.

The bottom panel of figure \ref{fig:deltasig_early_late} shows the inversion of
\deltasig\ to \xigm\ for each class.  The non-power law behavior for the early
types is still visible for \xigm\ but it is less pronounced.  The strong low
feature in the blue/late samples is related to the feature at smaller radius in
\deltasig; recall that the inversion to \xigm\ involves the derivative of
\deltasig.  

We fit a power law \xigm\ to each sample, using the full covariance matrix for
the red/early samples, but only the diagonal elements for the blue/late samples
because their covariance matrix is quite noisy.  The results are shown in
Table \ref{tab:allsamp}.  The red/early type galaxies have a higher amplitude
than the blue/late types, as indicated by the best-fit correlation length
$r_0$. The red/early subsamples are also nearly twice as luminous as the
blue/late subsamples.  We detect no appreciable difference in the slope of the
blue/late correlation function relative to that of the red/early types, in
contrast to the strong shift in slope of the galaxy auto-correlation function
as a function of type and color \citep{Zehavi02}. On the other hand, the errors
in the blue/late sample slopes are sizable, reflecting the lower S/N.

\subsection{Dependence of \xigm\ on Luminosity for Red Galaxies} 
\label{gmcfredlum}

\begin{figure*}[p]
\plotone{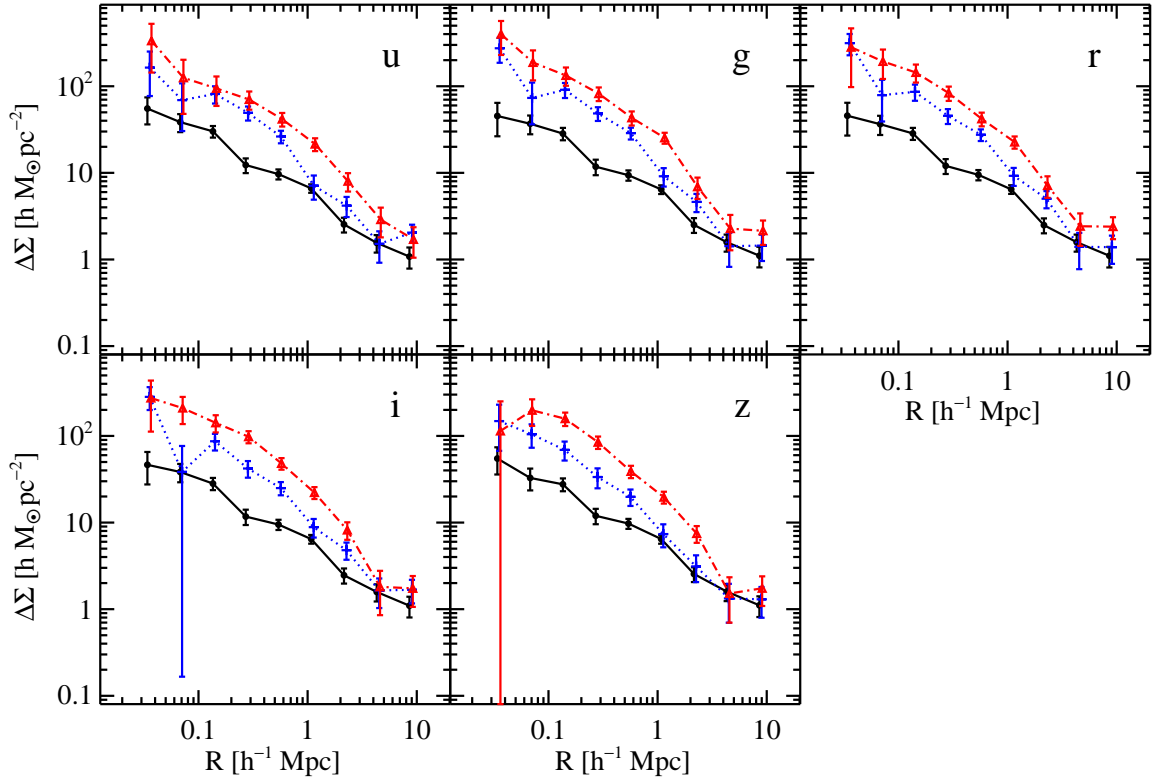} \figcaption{ 
Same as figure \ref{fig:deltasig_allband_bylum} but only for red galaxies, 
\gmr\ $ > $ \gmrcut.
\label{fig:deltasig_red_allband_bylum} }
\end{figure*}

\begin{figure*}[p]
\plotone{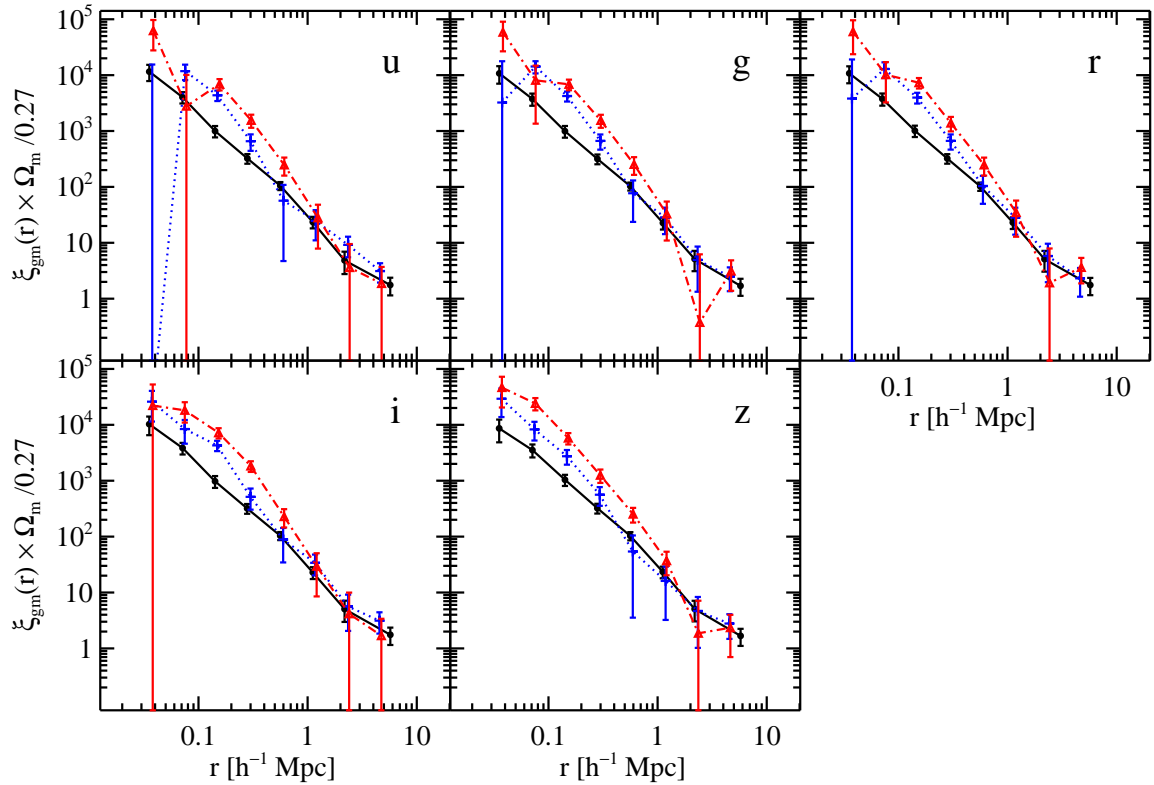} \figcaption{ 
Same as figure \ref{fig:xi_allband_bylum} but only for red galaxies, 
\gmr\ $ > $ \gmrcut.
\label{fig:xi_red_allband_bylum} }
\end{figure*}

The lensing S/N for the red galaxy sample is quite high, allowing us to further
subdivide this sample into three bins of luminosity in each of the 5 SDSS
bandpasses, as we did in \S \ref{gmcflum}.  The split is again 85\%, 10\%, and
5\% of the galaxies in each of the three luminosity bins.  The mean \deltasig\
for these luminosity bins is shown in Figure
\ref{fig:deltasig_red_allband_bylum}, and the inversion to \xigm\ is shown in
figure \ref{fig:xi_red_allband_bylum}.  The trends of \deltasig\ and \xigm\
with red galaxy luminosity are similar to those for the full sample.  The fits
to power law correlation functions are listed in Table \ref{tab:earlylumbin}.
Note that the correlation length $r_0$ and mean luminosity for each red
luminosity subsample is larger than that of the corresponding full luminosity
subsample, as expected from \S \ref{gmcftype}.

We now have two pieces of evidence suggesting that the increase in clustering
strength and change in shape we observe is more closely associated with
luminosity than with color: (i) for the subsamples split only by luminosity
(not color), galaxies split by \umag-band luminosity show a significant
increase of clustering strength with luminosity, and a change in shape, even
though the mean $g-r$ color is nearly independent of luminosity for these
subsamples; (ii) within the red galaxy sample itself, we find the same 
trends that we see for the full sample.

\subsection{Dependence of \xigm\ on Velocity Dispersion}  \label{gmcfvdis}

\begin{figure}[b]
\plotone{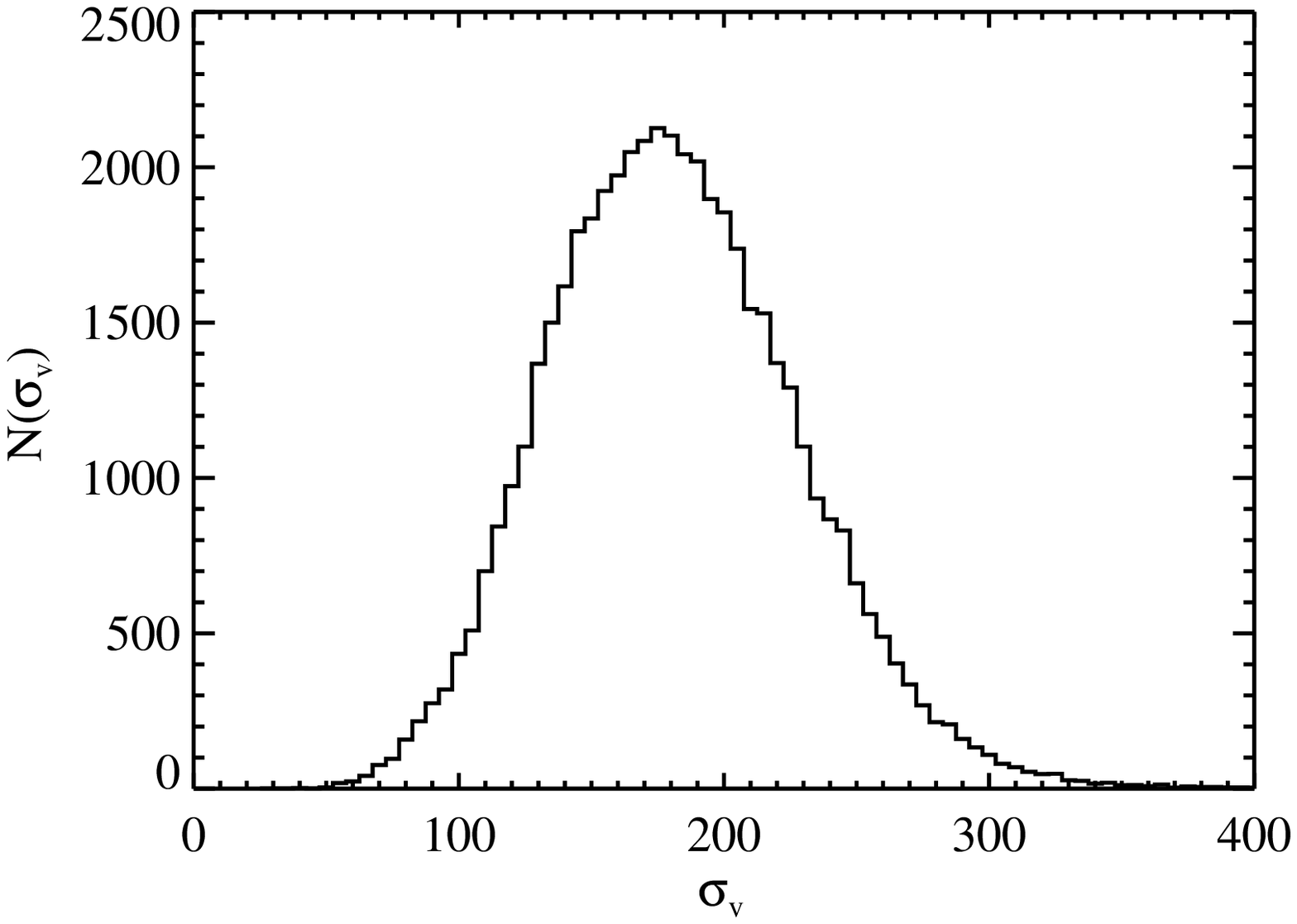} \figcaption{
Histogram of measured velocity
dispersion $\sigma_v$ for \numvdis\ early-type lens galaxies.
\label{fig:vel_dis_hist}}
\end{figure}

For a large fraction of the early type galaxies, selected by \eclass\ and by
the requirement that the surface brightness approximately fits a de Vaucouleurs
profile, the SDSS spectroscopic pipeline measures the velocity dispersion
$\sigma_v$. The distribution of velocity dispersions is shown in figure
\ref{fig:vel_dis_hist}.  We use only galaxies with $50 < \sigma_v ({\rm
km/sec}) < 400$ for this study and split the sample roughly in half at
$\sigma_v = 182$ km/sec.  The low-$\sigma_v$ sample has a mean of $\sigma_v =
143$ km/s and the high-$\sigma_v$ sample has a mean of $\sigma_v = 215$ km/s.

\begin{figure}[ht]
\plotone{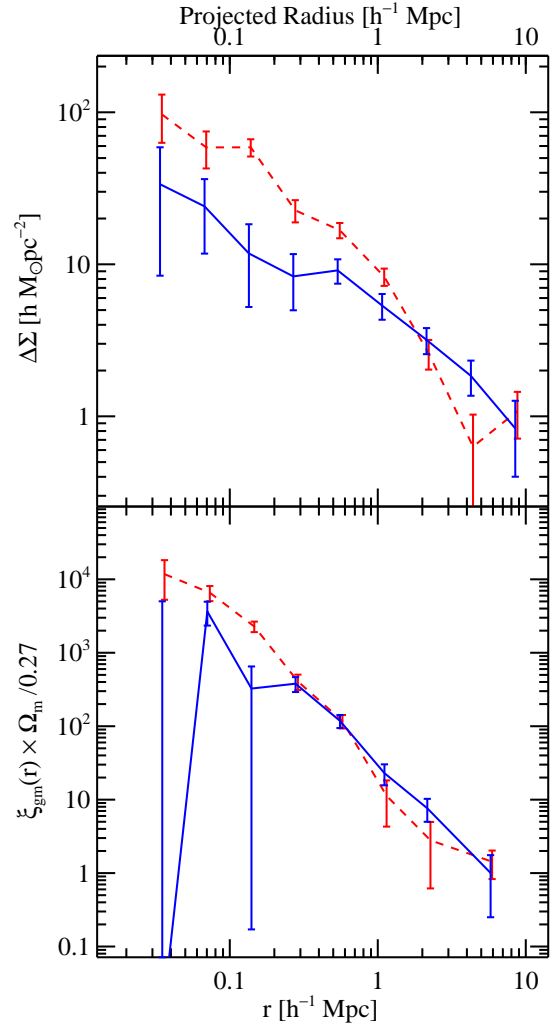} \figcaption{ Mean \deltasig\ and \xigm\
for early type galaxies with measured velocity dispersion $\sigma_v > 182$
km/sec \vdisred\ and $< 182$ km/sec \vdisblue.  The amplitude
for high $\sigma_v$ galaxies is stronger at small scales, as expected from the
trend with luminosity.  
\label{fig:deltasig_vdis} }
\end{figure}

In Figure \ref{fig:deltasig_vdis} we show \deltasig\ and \xigm\ for these two
samples of early-type galaxies split by velocity dispersion.  The half of the
sample with higher $\sigma_v$ shows a steeper slope, and a larger amplitude at
$r < 500 h^{-1}$ kpc, than the low-$\sigma_v$ galaxies. Interestingly, the
amplitude for the low $\sigma_v$ galaxies is actually somewhat higher that that
for high-$\sigma_v$ galaxies for $r > 1$Mpc.  For neither sample is \xigm\ well
fit by a power law.  The shape of the high-$\sigma_v$ \xigm\ is similar to that
for the high luminosity red subsample.  Given the Faber-Jackson relation
between velocity dispersion and luminosity, this trend agrees qualitatively
with the results above for the luminosity-scaling of \xigm\ for red galaxies.

\section{Discussion} \label{discussion}

We have presented measurements of the mean lensing signal $\Delta\Sigma =
\overline{\Sigma}(<R) - \overline{\Sigma}(R)$ for a flux limited spectroscopic
sample of SDSS galaxies with $\langle L \rangle \sim L_*$ and $\langle z
\rangle \simeq 0.1$ over scales 25$h^{-1}$ kpc to 10$h^{-1}$ Mpc.  \deltasig\
is consistent with a power-law for all separations, with index $0.76 \pm 0.05$.

We performed a direct inversion to the three-dimensional galaxy-mass
cross-correlation function \xigm\ over scales 
$0.025-6.7 h^{-1}$ Mpc\footnote[2]{Modelers should predict \deltasig\ if
possible. It is higher S/N and contains more information than the
inversion.}. The resulting \xigm\ is consistent with a power-law galaxy-mass
correlation function over this range, $\xi_{gm}\times (\Omega_m/0.27) =
(r/r_0)^{-\gamma}$.  The measured slope, $\gamma =1.79 \pm 0.06$ is consistent
with the slope of the auto-correlation function $\xi_{gg}$ for a similarly
selected set of galaxies \citep{Zehavi03}.  We find that the measured scale
length, $r_0 = (5.4 \pm 0.7)(0.27/\Omega_m)^{1/\gamma}$, is also consistent
with that of \xigg\ for the fiducial value of $\Omega_m = 0.27$, although the
range of radii probed by the two data sets is quite different: the galaxy
clustering results are measured for $r > 0.2 h^{-1}$ Mpc.

We compared \xigm\ with \xigg\ from \citet{Zehavi03} over the common range of
radii: $0.2-6.7 h^{-1}$ Mpc.  The ratio of these functions is
$\xi_{gg}/\xi_{gm} = b/r$, where $b$ is the bias and $r$ is the correlation
coefficient at a fixed separation.  We find that $b/r$ is consistent with being
scale independent over this range of scales, with a mean value of $\langle b/r
\rangle = (1.3 \pm 0.2)(\Omega_m/0.27)$. Semi-analytic galaxy formation models
suggest that $b/r$ is a direct measure of the standard bias $b$, i.e., that $r$
approaches unity for separations larger than 1 Mpc \citep{Guzik01}, although
there are hints from other lensing data that this may not be the case
\citep{Hoekstra02b}.

We repeated this analysis for a volume and magnitude limited sample of
\numspecvlim\ luminous galaxies, with selection criteria designed to match that
of the most luminous sample of \citet{Zehavi02}: $0.1 < z < 0.174$, $-23.0 <
M_r -5 \log(h) < -21.5$.  We detect no scale dependence in $b/r$, although
\xigm\ for this sample is noisier than the overall sample.  We find a mean
$b/r$ of $(2.0 \pm 0.7)(\Omega_m/0.27)$ over the common range of scales
($0.4-6.7 h^{-1}$ Mpc).

These results are consistent with and extend previous galaxy lensing
measurements.  \citet{fis00} found a power law \deltasig\ for angular scales
$\theta < 600$\arcsec, or roughly $R < 1 h^{-1}$ Mpc.  \citet{Wilson01} and
\citet{Hoekstra03a} (which incorporates some lens redshifts) found similar
results over a smaller range of scales, although the power law index was less
well constrained. The high precision measurements of \citet{Hoekstra03b} show a
power law on $\sim$2 Mpc scales (no power law fit is quoted).  \citet{Mckay02},
which used a smaller sample of SDSS lens galaxies with known redshifts, found a
power law \deltasig\ with slope consistent with the current study for $0.025 <
R < 1 h^{-1}$ Mpc.  An model-dependent SDSS lensing analysis concentrating on
small scales \citep[in preparation]{Seljak04} found results consistent with
those presented here.  \citet{Hoekstra02b} used a deeper, narrower sample than
ours, without lens redshifts, to obtain a high S/N measurement of the shear
over effective scales $R< 5 h^{-1}$ Mpc.  They find that $b/r$ is approximately
scale-dependent over the range $0.1-3 h^{-1}$ Mpc, with mean amplitude $b/r =
1.09 \pm 0.04$ over these scales. However, combining their galaxy lensing data
with deeper cosmic shear data, they measured both $b$ and $r$.  The errors in
$b$ and $r$ are correlated, since both depend on the cosmic shear, but they
found that $b$ and $r$ each vary with scale and track one another in such a way
that the ratio $b/r$ is approximately constant.

The value of $b/r$ inferred above requires only the following assumptions: that
general relativity is valid, the cosmological principle holds, and the FRW
metric properly relates redshift to distance in our universe.  We have
explicitly shown the dependence on the parameter $\Omega_m$ (the dependence on
other parameters through the angular diameter distance is weak for these
redshifts).  Further interpretation of these measurements requires further
assumptions and model dependence.  For example, if one believes that $L_*$
galaxies are nearly unbiased tracers of the mass on these scales, say $b/r =
b_{lin} < 2$, then one can place a limit $\Omega_m < 0.6$ at 95\%
confidence. We leave it to the reader to further interpret these measurements
in the context of particular models.

Splitting our lens galaxies into subsamples has enabled us to study $\xi_{gm}$
as a function of luminosity, color, and spectral type.  The amplitude of
$\xi_{gm}$ increases with increasing luminosity on intermediate ($0.1-1 h^{-1}$
Mpc) scales, consistent with the lower S/N results of \citet{Mckay02}. However,
we find that \xigm\ is nearly independent of luminosity on scales larger than
$1 h^{-1}$ Mpc (see figure \ref{fig:xi_allband_bylum}). This is indicative of
the increase in slope with luminosity on intermediate scales.  The highest
luminosity bin shows significant deviation from a power law, with a flattening
at radii less than $0.1 h^{-1}$ Mpc.  These trends of the slope and amplitude
of $\xi_{gm}$ with luminosity are in good qualitative agreement with the
predictions of semi-analytic galaxy formation models \citep{Guzik01}.  The
semi-analytic models predict that the amplitude of $\xi_{gm}$ increases with
luminosity on small scales, but is only weakly dependent on luminosity for
scales $r > 1 h^{-1}$ Mpc, in agreement with our results.

The increase in clustering amplitude with luminosity on intermediate scales for
\xigm\ is similar to that seen for the galaxy-autocorrelation function
\citep{Norberg01,Zehavi02}, but the deviation from a power law on small scales
at higher luminosity has not been seen in previous autocorrelation
measurements.  Also, the auto-correlation function shows a nearly equal
increase in amplitude with luminosity on both large and intermediate
scales \citep{Norberg01, Zehavi02}, but \xigm\ does not. This may imply a
scale-dependent bias for high luminosities.

We have also detected a trend in $\xi_{gm}$ with galaxy color and spectral type
(figure \ref{fig:deltasig_early_late}). Red/early type galaxies show stronger
correlation amplitude than blue/late types but the slopes of $\xi_{gm}$ for the
two subsamples are consistent within the errors. These results are consistent
with the previous results of \citet{Mckay02}. This behavior is also in
qualitative agreement with the semi-analytic galaxy formation predictions
\citep{Guzik01}.

Finally, we have sufficient S/N to study the scaling of $\xi_{gm}$ with
luminosity and velocity dispersion for red/early type galaxies.  Again, we find
a similar increase in logarithmic slope and amplitude with luminosity on
intermediate scales, and a weak luminosity dependence on large scales. We find
the slope of \xigm\ increases with $\sigma_v$, but the high $\sigma_v$ sample
shows significant deviations from a power law.

A separate analysis will compare these results in detail with predictions from
N-body simulations and the halo model of structure formation.  A future study,
using a larger lens sample, will concentrate on galaxy-galaxy lensing at small
separations, where the signal is primarily sensitive to galaxy dark matter
halos.


\acknowledgments


We thank Andreas Berlind for many useful discussions.

ES and DEJ are supported by the CFCP under NSF PHY 0114422; JAF by the CFCP and
the DOE, NASA grant NAG5-10842; RS by ITR 01211671; TAM by NSF AST 9703282
and AST 0206277; and AJC by AST 9984924 and LTSA NAG58546.  Funding for the
creation and distribution of the SDSS Archive has been provided by the Alfred
P. Sloan Foundation, the Participating Institutions, the National Aeronautics
and Space Administration, the National Science Foundation, the U.S. Department
of Energy, the Japanese Monbukagakusho, and the Max Planck Society. The SDSS
Web site is http://www.sdss.org/.  The SDSS is managed by the Astrophysical
Research Consortium (ARC) for the Participating Institutions. The Participating
Institutions are the University of Chicago, Fermilab, the Institute for
Advanced Study, the Japan Participation Group, the Johns Hopkins University,
Los Alamos National Laboratory, the Max-Planck-Institute for Astronomy (MPIA),
the Max-Planck-Institute for Astrophysics (MPA), New Mexico State University,
the University of Pittsburgh, Princeton University, the United States Naval
Observatory, and the University of Washington.

\section{Appendix: Photoz Systematic Errors and Intrinsic Alignments} \label{systematics}

We discussed two sources of systematic errors in \S \ref{gglmeas:random}:
residuals from the PSF correction and clustering of faint sources around the
lens galaxies.  These biases have been estimated from the data and corrected
for in our final results.

Another possible source of error may come from biases in the photometric
redshifts (\S \ref{data:imaging:photoz}).  Although the photometric redshift
distributions inferred for SDSS galaxies are consistent with published redshift
surveys for \rmag$~<21$ \citep{Csabai2003}, there are well-known degeneracies
in the technique (between type and redshift) that tend to push galaxies to
particular values of photoz.  This pile-up can be seen in figure
\ref{fig:photozdist}: there are significant peaks in the photoz distribution at
several redshifts. Although such peaks are often seen in spectroscopic redshift
surveys, where they are associated with large-scale structures, the volume
probed by the SDSS photoz sample is so large, containing $\sim 4$ million
galaxies covering $\sim 1000$ square degrees out to $z \sim 0.6$, that the
effects of such structures should be very small.

Because galaxies in parameter regions with strong degeneracies have relatively
large photoz errors, we expect this bias to be suppressed in the final
analysis. We have repeated our analysis using only the source galaxy redshift
prior inferred from summing the individual Gaussians (as shown in figure
\ref{fig:photozdist}), i.e., not using the photoz measurements for individual
source galaxies, and we recover the same results within the noise.  We have
also used the method outlined in \citet{Mckay02}, which uses the \rmag\
magnitudes to infer the redshift distribution, and recover consistent results.
Thus the bias from photoz degeneracies appears to be small compared to our
statistical uncertainty.


Another issue that can possibly complicate interpretation of our results is
intrinsic correlations in the ellipticities of the source galaxies.  If
the orientations of physically associated pairs of galaxies are correlated,
this could possibly mimic a lensing signal.

Tidal torques can cause the angular momentum vectors of physically associated
galaxies to be aligned \citep{Critt01}. The magnitude of these ellipticity
correlations is expected to be strong and relatively constant on scales less
than $\sim$\ 1 Mpc, and decrease as $\xi_{LS}$, the correlation function of
sources with lenses, on large scales \citep{Critt01}.  This effect has been
seen in N-body simulations \citep{CroftMetzler00,Heavens00} and there are
possible detections in the Tully catalog \citep{Pen00} and SuperCOSMOS Sky
Survey data \citep{Brown02}, although these data are not ideal for such a
study.  The recent work of \citet{Heymans03} suggests this effect may be near
the low end of theoretical predictions.

It is not clear how intrinsic alignments affect tangential shear measurements.
The tangential shear is the correlation between the position of a lens and the
projection of source ellipticities onto the tangential reference frame, defined
by a circle centered at the lens. The intrisic ellipticity correlations
described above are correlations between pairs of ellipticities.  There is
little theoretical guidance about what tangential alignment to expect.  For
limits on this effect we turn to the recent study by \citet{BernNorb02} which
uses data from the 2dF redshift survey to study the tangential alignment of
faint (2.2 magnitudes fainter than the primary) physically associated
galaxies. They place an upper limit of $\langle e_+^{int} \rangle < 0.02$(95\%)
within 500 $h^{-1}$ kpc at $\langle z \rangle = 0.1$. A recent analysis of this
tangential alignment within the SDSS spectroscopic sample \citep{Koester03}
places still more stringent limits on this tangential alignment $\langle
e_+^{int} \rangle < 0.007 $(95\%) on $500 h^{-1}$ kpc scales.

Most of the sources used in our study are far behind the lenses, in which case
no correlation is expected between lens position and the tangential source
galaxy ellipticity. A small fraction of sources, however, are actually faint
companions of the foreground lenses.  The fraction of physically associated
sources is generally small \citep{fis00}, and the use of photometric redshifts
in our study down-weights these sources in the final analysis.  The relative
contribution of these objects in the lensing measurement is shown in figure
\ref{fig:cluster_corr}. The mean contribution over 500 kpc is 0.018.

Assuming the result from \citet{BernNorb02} also applies to our faint sources,
we place a limit on the contamination from intrinsic alignments at $\langle
e_+^{int} \rangle < 0.02*0.018 = 3.6\times 10^{-4}$ on 500$ h^{-1}$ kpc
scales. The mean distortion within this radius is $1.5 \pm 0.1 \times
10^{-3}$. Thus, on 500 $h^{-1}$ kpc scales, the intrinsic alignments should
contribute less than 20\% of the signal at 95\% confidence. The new SDSS
results suggest the effect is smaller still ($< 8.5\%$). The intrinsic
alignment correlations decrease with scale roughly as $\xi_{LS}$, and the
relative contribution from the sources also decreases quickly with scale
(figure \ref{fig:cluster_corr}). Thus, the contribution from intrinsic
alignments should be small for radii $r \gtrsim 500 h^{-1}$ kpc.


\newpage



\newpage

\newpage


\begin{deluxetable}{cccccccc}
\tabletypesize{\small}
\tablecaption{Model Fits for Different Lens Samples\label{tab:allsamp}}
\tablewidth{0pt}
\tablehead{
\colhead{Sample} &
\colhead{Selection Criteria} &
\colhead{Mean Abs. Mag.} &
\colhead{Mean \gmr} &
\colhead{N$_{Lenses}$} &
\colhead{$r_0$} &
\colhead{$\gamma$} &
\colhead{$\chi^2/\nu$} 
}
\
\startdata
 All & - & -20.767 (1.455 $\pm$ 0.004) & 0.629 & 127001 & 5.4 $\pm$ 0.7 & 1.79 $\pm$ 0.06 & 17.1/15 \\
 Red & \gmr\ $ > $ \gmrcut\ & -21.061 (1.908 $\pm$ 0.006) & 0.753 & 60099 & 6.9 $\pm$ 0.8 & 1.81 $\pm$ 0.05 & 12.3/15 \\
 Blue & \gmr\  $ < $ \gmrcut\ & -20.477 (1.114 $\pm$ 0.004) & 0.536 & 65134 & 4.0 $\pm$ 1.0 & 1.76 $\pm$ 0.15 & 5.97/6 \\
 Early & \eclass\ $ < $ \eclasscut\ & -21.036 (1.864 $\pm$ 0.006) & 0.737 & 62340 & 7.3 $\pm$ 0.8 & 1.77 $\pm$ 0.05 & 16.9/15 \\
 Late & \eclass\ $ > $ \eclasscut\ & -20.474 (1.111 $\pm$ 0.004) & 0.539 & 64378 & 3.3 $\pm$ 1.0 & 1.89 $\pm$ 0.17 & 7.18/6 \\
 Vlim & $ -23.0 < M_r < -21.5 $ & -21.854 (3.961 $\pm$ 0.012) & 0.718 & 10277 & 5.1 $\pm$ 0.8 & 2.01 $\pm$ 0.07 & 24.6/15 \\
   & $   0.1 < z   < 0.174 $ & & & & & & 
\enddata
\tablecomments{Absolute magnitudes are \rmag-band Petrosian $M - 5 \log_{10} h$. 
               Values in parentheses are luminosity in units of $10^{10} h^{-2} L_{\sun}$.
               The means are calculated using the same weights as the lensing measurement.
               The value of $M_* (L_*)$ for the \rmag-band is -20.83 (1.54).
               The $r_0$ and $\gamma$ are best fit parameters for $\xi_{gm} = 
               (r/r_0)^{-\gamma}$; $r_0$ is measured in $h^{-1}$ Mpc. A value of
               $\Omega_m = 0.27$ was assumed. 
	       For the ``late'' and ``blue'' samples, the data were rebinned from 17 to 8 radial bins.}
\end{deluxetable}

\begin{deluxetable}{cccccccc}
\tabletypesize{\small}
\tablecaption{Luminosity Bins for All Galaxies \label{tab:lumbin}}
\tablewidth{0pt}
\tablehead{
\colhead{Bandpass} &
\colhead{Abs. Mag. Range} &
\colhead{Mean Abs. Mag.} &
\colhead{Mean \gmr} &
\colhead{N$_{Lenses}$} &
\colhead{$r_0$} &
\colhead{$\gamma$} &
\colhead{$\chi^2/\nu$} 
}
\
\startdata
$u$ & -19.6 $ < M_{u} < $ -15.0 & -18.515 (0.908 $\pm$ 0.002) & 0.628 & 106585 & 5.5 $\pm$ 0.8 & 1.75 $\pm$ 0.06 & 13.5/15 \\
 -  & -20.0 $ < M_{u} < $ -19.6 & -19.763 (2.864 $\pm$ 0.003) & 0.639 & 12539 & 4.4 $\pm$ 0.8 & 2.04 $\pm$ 0.09 & 19.2/15 \\
 -  & -22.0 $ < M_{u} < $ -20.0 & -20.318 (4.779 $\pm$ 0.020) & 0.654 & 6270 & 6.9 $\pm$ 1.0 & 1.97 $\pm$ 0.07 & 26.3/15 \\
    & & & & & & \\
$g$ & -21.0 $ < M_{g} < $ -16.5 & -19.891 (0.956 $\pm$ 0.002) & 0.623 & 106646 & 5.4 $\pm$ 0.9 & 1.75 $\pm$ 0.07 & 15.1/15 \\
 -  & -21.4 $ < M_{g} < $ -21.0 & -21.188 (3.156 $\pm$ 0.004) & 0.699 & 12546 & 5.0 $\pm$ 0.8 & 2.02 $\pm$ 0.08 & 12.7/15 \\
 -  & -23.5 $ < M_{g} < $ -21.4 & -21.718 (5.145 $\pm$ 0.025) & 0.734 & 6275 & 7.0 $\pm$ 0.8 & 2.03 $\pm$ 0.06 & 24.6/15 \\
    & & & & & & \\
$r$ & -21.7 $ < M_{r} < $ -17.0 & -20.544 (1.184 $\pm$ 0.003) & 0.621 & 106643 & 5.4 $\pm$ 0.8 & 1.74 $\pm$ 0.07 & 15.9/15 \\
 -  & -22.2 $ < M_{r} < $ -21.7 & -21.902 (4.137 $\pm$ 0.005) & 0.726 & 12543 & 5.2 $\pm$ 0.8 & 2.03 $\pm$ 0.07 & 13.5/15 \\
 -  & -24.0 $ < M_{r} < $ -22.2 & -22.463 (6.937 $\pm$ 0.026) & 0.767 & 6274 & 7.5 $\pm$ 0.8 & 2.01 $\pm$ 0.05 & 30.9/15 \\
    & & & & & & \\
$i$ & -22.0 $ < M_{i} < $ -17.0 & -20.870 (1.446 $\pm$ 0.003) & 0.620 & 106625 & 5.3 $\pm$ 0.9 & 1.74 $\pm$ 0.07 & 17.6/15 \\
 -  & -22.5 $ < M_{i} < $ -22.0 & -22.232 (5.067 $\pm$ 0.006) & 0.734 & 12544 & 5.1 $\pm$ 0.7 & 2.04 $\pm$ 0.07 & 19.3/15 \\
 -  & -24.0 $ < M_{i} < $ -22.5 & -22.776 (8.368 $\pm$ 0.030) & 0.772 & 6271 & 6.9 $\pm$ 0.8 & 2.06 $\pm$ 0.06 & 24.7/15 \\
    & & & & & & \\
$z$ & -22.2 $ < M_{z} < $ -17.0 & -21.069 (1.721 $\pm$ 0.004) & 0.619 & 105975 & 5.3 $\pm$ 0.9 & 1.73 $\pm$ 0.07 & 16.8/15 \\
 -  & -22.6 $ < M_{z} < $ -22.2 & -22.350 (5.599 $\pm$ 0.007) & 0.725 & 12466 & 5.5 $\pm$ 0.8 & 1.97 $\pm$ 0.07 & 20.3/15 \\
 -  & -24.0 $ < M_{z} < $ -22.6 & -22.881 (9.132 $\pm$ 0.032) & 0.753 & 6236 & 6.6 $\pm$ 0.8 & 2.05 $\pm$ 0.06 & 22.1/15
\enddata
\tablecomments{Galaxies were split into three bins of absolute magnitude in 
     each of the five SDSS bandpasses, as listed in column two. See table 
     \ref{tab:allsamp} for explanations of the other columns. The 
     value of $M_* (L_*)$ is -18.34(0.77), -20.04(1.10), 
     -20.83(1.54), -21.26(2.07), -21.55(2.68) for $u, g, r, i, z$ 
     respectively.}
\end{deluxetable}

\begin{deluxetable}{cccccccc}
\tabletypesize{\small}
\tablecaption{Luminosity Bins for Red Galaxies \label{tab:earlylumbin}}
\tablewidth{0pt}
\tablehead{
\colhead{Bandpass} &
\colhead{Abs. Mag. Range} &
\colhead{Mean Abs. Mag.} &
\colhead{Mean \gmr} &
\colhead{N$_{Lenses}$} &
\colhead{$r_0$} &
\colhead{$\gamma$} &
\colhead{$\chi^2/\nu$} 
}
\
\startdata
$u$ & -19.6 $ < M_{u} < $ -15.0 & -18.521 (0.913 $\pm$ 0.003) & 0.752 & 51074 & 6.9 $\pm$ 0.9 & 1.77 $\pm$ 0.06 & 10.0/15 \\
 -  & -20.1 $ < M_{u} < $ -19.6 & -19.791 (2.939 $\pm$ 0.005) & 0.769 & 6008 & 5.6 $\pm$ 0.8 & 2.09 $\pm$ 0.07 & 21.6/15 \\
 -  & -22.0 $ < M_{u} < $ -20.1 & -20.373 (5.026 $\pm$ 0.027) & 0.789 & 3006 & 9.6 $\pm$ 1.1 & 1.96 $\pm$ 0.06 & 35.2/15 \\
    & & & & & & \\
$g$ & -21.2 $ < M_{g} < $ -16.5 & -20.112 (1.171 $\pm$ 0.003) & 0.752 & 51084 & 7.0 $\pm$ 1.0 & 1.75 $\pm$ 0.06 & 9.46/15 \\
 -  & -21.6 $ < M_{g} < $ -21.2 & -21.364 (3.712 $\pm$ 0.006) & 0.773 & 6008 & 6.4 $\pm$ 0.7 & 2.04 $\pm$ 0.06 & 20.8/15 \\
 -  & -23.5 $ < M_{g} < $ -21.6 & -21.841 (5.762 $\pm$ 0.025) & 0.788 & 3006 & 8.9 $\pm$ 0.9 & 2.03 $\pm$ 0.05 & 40.2/15 \\
    & & & & & & \\
$r$ & -21.9 $ < M_{r} < $ -17.0 & -20.867 (1.596 $\pm$ 0.005) & 0.751 & 51080 & 7.0 $\pm$ 1.0 & 1.75 $\pm$ 0.06 & 9.56/15 \\
 -  & -22.4 $ < M_{r} < $ -21.9 & -22.140 (5.152 $\pm$ 0.009) & 0.775 & 6008 & 6.6 $\pm$ 0.8 & 2.03 $\pm$ 0.06 & 19.0/15 \\
 -  & -24.0 $ < M_{r} < $ -22.4 & -22.635 (8.127 $\pm$ 0.037) & 0.793 & 3004 & 9.0 $\pm$ 0.9 & 2.03 $\pm$ 0.05 & 34.3/15 \\
    & & & & & & \\
$i$ & -22.3 $ < M_{i} < $ -17.0 & -21.240 (2.033 $\pm$ 0.006) & 0.751 & 51078 & 6.9 $\pm$ 1.0 & 1.76 $\pm$ 0.06 & 10.1/15 \\
 -  & -22.7 $ < M_{i} < $ -22.3 & -22.478 (6.354 $\pm$ 0.011) & 0.776 & 6007 & 7.1 $\pm$ 0.9 & 1.98 $\pm$ 0.06 & 20.9/15 \\
 -  & -24.0 $ < M_{i} < $ -22.7 & -22.946 (9.778 $\pm$ 0.043) & 0.790 & 3005 & 8.4 $\pm$ 0.8 & 2.06 $\pm$ 0.05 & 35.0/15 \\
    & & & & & & \\
$z$ & -22.4 $ < M_{z} < $ -17.0 & -21.412 (2.360 $\pm$ 0.007) & 0.751 & 50697 & 7.0 $\pm$ 1.0 & 1.75 $\pm$ 0.06 & 10.1/15 \\
 -  & -22.8 $ < M_{z} < $ -22.4 & -22.563 (6.810 $\pm$ 0.011) & 0.772 & 5962 & 5.4 $\pm$ 0.9 & 2.06 $\pm$ 0.08 & 8.99/15 \\
 -  & -24.0 $ < M_{z} < $ -22.8 & -23.045 (10.61 $\pm$ 0.048) & 0.782 & 2983 & 7.2 $\pm$ 0.9 & 2.09 $\pm$ 0.06 & 26.0/15
\enddata
\tablecomments{Same as table \ref{tab:lumbin} but for red galaxies, defined
               as galaxies with \gmr\ $ > 0.7$.}
\end{deluxetable}

\end{document}